\documentclass[twocolumn, pt12]{aastex63}

\usepackage[T1]{fontenc}
\usepackage{graphicx}	% Including figure files
\usepackage{amsmath}	% Advanced maths commands
\usepackage{amssymb}	% Extra maths symbols
\usepackage{hyperref}
\usepackage{makecell}
\usepackage{color} 		% Colors for text commenting
\usepackage{comment}    % Allows for commenting out blocks of text

\newcommand{\R} {\text{RMS}} 
\newcommand{\thetaR} {\theta_\text{RMS}}
\newcommand{\thetaC} {\theta_\text{C}}
\newcommand{\thetaU} {\theta_\text{u}}
\newcommand{\meanEpsilon} {\left< \epsilon \right>}
\newcommand{\us} {\text{u}}
\newcommand{\ds} {\text{d}}
\newcommand{\rbar} {\bar{r}}
\newcommand{\Rph} {R_\text{ph}}

\newcommand{\radshock} {\texttt{radshock}}
\newcommand{\komrad} {\texttt{Komrad}}

\def\cl#1{{#1}}

\shortauthors{Samuelsson, Lundman, Ryde}
\shorttitle{The Kompaneets RMS approximation}

\begin{document}
\label{firstpage}
%\title{A Fokker-Planck solution to radiation mediated shocks in photon rich environments}
%\title{Radiation mediated shock model for the prompt emission of gamma-ray bursts}
%\title{\cl{The Kompaneets radiation-mediated-shock approximation:\\ How to fit radiation-mediated-shock emission to prompt gamma-ray burst data}}
\title{An efficient method for fitting radiation-mediated shocks to gamma-ray burst data:\\The Kompaneets RMS approximation}
%\title{\cl{How to fit radiation-mediated shock emission to prompt gamma-ray burst data}}

%%%%%%%%%%%%%%%%%%%%%%%%%%%%%%%%%%%%%%%%%%%%%%%%%%%%%%%%%%%%%%%
% Layout APJ aastex6.3

%%%%%%%%%%%%%%%%%%%%%%%%%%%%%%%%%%
%%%%%%%%%%%%%% AUTHORS %%%%%%%%%%%%%%
%%%%%%%%%%%%%%%%%%%%%%%%%%%%%%%%%%
\correspondingauthor{Filip Samuelsson}
\email{filipsam@kth.se}

\author[0000-0001-7414-5884]{Filip Samuelsson}
\affiliation{Department of Physics, KTH Royal Institute of Technology, \\
and The Oskar Klein Centre, SE-10691 Stockholm, Sweden}

\author[0000-0002-0642-1055]{Christoffer Lundman}
\affiliation{The Oskar Klein Centre, Department of Astronomy, \\ 
Stockholm University, AlbaNova, SE-10691 Stockholm, Sweden}

\author[0000-0002-9769-8016]{Felix Ryde}
\affiliation{Department of Physics, KTH Royal Institute of Technology, \\
and The Oskar Klein Centre, SE-10691 Stockholm, Sweden}

\begin{abstract}
%A promising candidate to produce the prompt emission of gamma-ray bursts (GRBs) are radiation-mediated shocks (RMSs) below the photosphere.
Shocks that occur below a gamma-ray burst (GRB) jet photosphere are mediated by radiation. Such radiation-mediated shocks (RMSs) could be responsible for shaping the prompt GRB emission.
Although well studied theoretically, RMS models have not yet been fitted to data due to the computational cost of simulating RMSs from first principles.
%Such shocks have recently been studied using first-principle radiation-hydrodynamics simulations. However, since these simulations are computationally expensive, no RMS model has yet been fitted against data. 
%Here, we develop a approximate method for simulating radiation spectra from mildly relativistic (in the shock frame) or slower RMSs, called the Kompaneets RMS approximation (KRA). %The KRA allows for fast simulations, which makes fitting to data possible. 
Here, we bridge the gap between theory and observations by developing an approximate method capable of accurately reproducing radiation spectra from mildly relativistic (in the shock frame) or slower RMSs, called the Kompaneets RMS approximation (KRA).
The approximation is based on the similarities between thermal Comptonization of radiation and the bulk Comptonization that occurs inside an RMS. We validate the method by comparing simulated KRA radiation spectra to first-principle radiation-hydrodynamics simulations, finding excellent agreement both inside the RMS and in the RMS downstream. The KRA is then applied to a shock scenario inside a GRB jet, allowing for fast and efficient fitting to GRB data. We illustrate the capabilities of the developed method by performing a fit to a non-thermal spectrum in GRB 150314A. The fit allows us to uncover the physical properties of the RMS responsible for the prompt emission, such as the shock speed and the upstream plasma temperature.% The spectrum is well fitted by a broad RMS spectrum with two breaks. From the fit we can thus deduce properties of the RMS, e.g. the amount of energy that must has been dissipated.
\end{abstract}
%%%%%%%%%%%%%%%%%%%%%%%%%%%%%%%%%%%%%%%%
%%%%%%%%%%%%% INTRODUCTION %%%%%%%%%%%%%
%%%%%%%%%%%%%%%%%%%%%%%%%%%%%%%%%%%%%%%%

\section{Introduction}

The launching, propagation and collimation of a highly supersonic jet unavoidably leads to immense shock formation inside the jet and its surroundings \citep[see e.g.,][]{Lopez-Camara2013, Lopez-Camara2014, Gottlieb2019}. Shocks that occur deep inside gamma-ray burst (GRB) jets are mediated by radiation. Such radiation-mediated shocks (RMSs) fill the jet with hot, non-thermal radiation, which is advected toward the jet photosphere where it is released. The released spectra will range from strongly non-thermal to thermal, depending on whether the radiation has had time to thermalize via scatterings before reaching the photosphere or not.

%Very narrow spectra that are undoubtedly thermal are rarely observed in the prompt emission of GRBs, which is often well described by the broader phenomenological Band function \citep{Band1993}. However, in a few instances where such narrow spectra have been observed, it has been seen followed by a gradual broadening of the spectrum within the same emission pulse \citep[see e.g.,][]{Ryde2011}. This clearly hints toward dissipation far below the photosphere \citep{ReesMeszaros2005}, and yet, so far, no RMS model has been fitted to the data.
GRBs are observed to have strong spectral evolution, both in terms of peak energy \citep{Golenetskii1983} and shape \citep[e.g., the width of the spectrum][]{Wheaton1973}. In around one quarter of GRB pulses, the narrowest, time resolved spectrum is consistent with a thermal spectrum, which strongly suggests that the whole pulse is of a photospheric origin \citep{Yu2019, Acuner2020, Dereli2020, Li2021}. It is plausible, therefore, that the wider, non-thermal spectra in such pulses have undergone subphotospheric dissipation \citep{ReesMeszaros2005, Ryde2011}. Even though RMSs are a natural cause of this dissipation, so far, RMS models have not been fitted to the data.\footnote{{We note that \citet{Ahlgren2015} and \citet{Vianello2018} fit photospheric models including dissipation to the data. However, their assumed energy dissipation mechanism was different.}%{See \citet{Ahlgren2015}, who did fit a photospheric model with dissipation to the data. However, their assumed energy dissipation mechanism was different and the fitted parameters could not be connected to the RMS.}
}

The main reason for this is that RMSs are expensive to simulate from first principles. RMSs have previously been considered in one spatial dimension \citep{LevinsonBromberg2008, NakarSari2012, Beloborodov2017, Lundman2018, Ito2018, LundmanBeloborodov2019, LevinsonNakar2020, Levinson2020, Ito2020, LundmanBeloborodov2021}. The 1D simulations illustrate the main features of the non-thermal RMS radiation expected inside GRB jets: a broad power-law spectrum for up to mildly relativistic shocks (that have relative relativistic speed between the up- and downstreams of $\beta\gamma \lesssim \,$few, where $\beta$ is the speed in units speed of light $c$ and $\gamma$ is the Lorentz factor), while faster shocks have more complex spectral shapes due to Klein-Nishina effects, anisotropic radiation in the shock, and photon-photon pair production. Once advected into the downstream, the RMS spectrum gradually thermalizes through scatterings.

Currently, these 1D simulations are not fast enough to build a table model of simulated RMS spectra over the relevant parameter space, which is an efficient way to test models against data. %Thus, GRB theories based on RMS radiation has so far not been properly tested against data for computational reasons. 
However, model testing is of crucial importance to further develop our understanding of the prompt emission in GRBs. With this motivation, in this work we explore an alternative path to connecting RMS theory and GRB observations. 
In Section~\ref{sec:KRA}, we construct an approximate, but very fast, method called the Kompaneets RMS approximation (KRA). The approximation is based on the strong similarities between bulk Comptonization of radiation inside an RMS, and thermal Comptonization of radiation on hot electrons, the latter being described by the Kompaneets equation. The KRA is appropriate to use for mildly relativistic (and slower), optically thick RMSs. We validate the KRA by comparing simulated radiation spectra to those produced by the full radiation hydrodynamics simulations, finding excellent agreement. The KRA is then applied to a minimal model of a shock inside a GRB jet in Section~\ref{Sec:jet_geometry}, which generate synthetic photospheric spectra, accounting for both adiabatic cooling and thermalization of the photon distribution. %which allowing for simulation of the (partially or fully thermalized) RMS spectrum that escapes the jet photosphere. 
The KRA simulations are about four orders of magnitude faster to run than the corresponding 1D simulations, allowing for table model construction. As an illustration of the model capabilities, we use the table model to perform a fit to the prompt emission of GRB 150314A in Section~\ref{Sec:Example_fit}. We conclude by summarizing and discussing our results in Section~\ref{Sec:Discussion}.

\section{The Kompaneets RMS approximation}
\label{sec:KRA}

In this section, we develop the KRA and compare the resulting spectra to full scale, special relativistic RMS simulations in planar geometry. The approximation is valid for RMSs where the photons inside the shock do not obtain energies exceeding the electron rest mass energy, as the transfer problem then becomes more complicated, including Klein-Nishina scattering effects, anisotropy, and $\gamma\gamma$-pair production. 
This typically corresponds to shocks that are mildly relativistic\footnote{We illustrate this point later with a shock simulation that has an upstream four-velocity of $\beta \gamma = 3$.} or slower, inside plasma where the downstream radiation pressure dominates over the magnetic pressure.

As is appropriate for RMSs inside GRB jets, the RMS is assumed to be photon rich \citep{Bromberg2011b}, i.e., the photons inside the RMS are mainly supplied by advection of upstream photons, as opposed to photon production inside the RMS and in the immediate downstream. The RMS is also assumed to be in an optically-thick region, which is appropriate deep below the photosphere. The approximation will therefore not hold for shocks that dissipate most of their energy close to the photosphere; such shocks require full radiation hydrodynamics simulations.\footnote{\citet{LundmanBeloborodov2021} shows the evolution of a mildly relativistic RMS that reaches the edge of neutron star merger ejecta. The shock evolution is complex: the radiation begins leaking ahead of the shock, while a forward and a reverse collisionless shock is formed at the photosphere.}

We note that \citet{BlandfordPayneII1981} showed that the shape of a photon spectrum traversing a photon rich, nonrelativistic RMS can be obtained analytically. Although their analytical calculation is impressive, it ignores photon energy losses due to electron recoil. The photon spectrum, therefore, lacks a high-energy cutoff, which makes it accurate only for soft spectra where the bulk energy is not carried by the high-energy photons. Due to this limitation, their solution is not applicable here.

\subsection{Bulk Comptonization inside the RMS}\label{Sec:energy_gain_RMS}
The following treatment assumes a nonrelativistic shock, but comparison to full RMS simulations show that the approximation is valid also for mildly relativistic shocks with $\gamma\beta \lesssim \,$few (see also Section \ref{subsec:estimating_kra_upper_speed_limit}).

In the shock rest frame, the incoming speed of the upstream is greater than the outgoing speed of the downstream, leading to a speed gradient inside the shock. The photons that diffuse inside the RMS speed gradient directly tap the incoming kinetic energy by scattering on fast electrons. If the photon mean free path is $\lambda$, the velocity difference of the plasma over a scattering length is $\lesssim \lambda(d\beta/dx) = d\beta/d\tau$, where $x$ is the spatial coordinate and $d\tau = dx/\lambda$ measures the optical depth along the $x$-coordinate. Doppler boosting the photon to the frame of the scatterer, performing a scattering and averaging over the scattering angles, one finds a relative energy gain of

\begin{equation}\label{eq:energy_gain_radshock}
    \frac{\Delta\epsilon}{\epsilon} \approx \frac{1}{3} \left| \frac{d\beta}{d\tau} \right|,
\end{equation}

\noindent per scattering, e.g., a first order Fermi process. Here, the photon energy $\epsilon$ is given in units of electron rest mass, $\epsilon = h\nu/m_{\rm e}c^2$, where $h$ is Planck's constant and $\nu$ is the photon frequency. Equation \eqref{eq:energy_gain_radshock} is valid for a relative energy gain $\Delta\epsilon/\epsilon < 1$, where $\Delta \epsilon$ is the energy gain in a scattering for a photon with initial energy $\epsilon$. Note that $d\beta/d\tau$ is a local quantity that changes continuously across the RMS transition region, and vanishes in the far up- and downstreams, where the plasma velocity is constant (see Figure \ref{Fig:Threezone} for a schematic of the velocity profile across the shock). The $d\beta/d\tau$-profile is self-consistently determined by the radiation feedback onto the plasma: the photons gain precisely the available kinetic energy such that the Rankine-Hugoniot shock jump conditions are satisfied.

Since plasma is advected through the RMS, so are the photons that scatter inside the plasma. However, photons also diffuse within the flow, and a fraction of the photons will stay inside the RMS much longer than the advection time across the RMS, accumulating more scatterings and therefore more energy. As is always the case when both the probability of escaping the shock and the relative energy gain per scattering, $\Delta\epsilon/\epsilon$, are energy independent, a power-law spectrum develops. The power-law extends up to energies where the energy gain per scattering is balanced by energy losses due to electron recoil, which occurs when $\Delta\epsilon/\epsilon \approx \epsilon$. This gives a maximum photon energy inside the shock of

\begin{equation}
    \epsilon_{\rm max} \approx \left<\frac{\Delta\epsilon}{\epsilon}\right>,
\label{eq:epsilon_max}
\end{equation}
where the brackets on the right-hand side indicates a weighted average across the shock. 

The exact expression for $\left< \Delta\epsilon/\epsilon \right>$ is difficult to determine from first principles. With $\bar{\epsilon}_{\rm u}$ and $\bar{\epsilon}_{\rm d}$ as the average photon energies in the up- and downstreams, respectively, and $\beta_{\rm u}\gamma_{\rm u}$ the four-velocity of the upstream evaluated in the shock rest frame, we empirically find in Appendix~\ref{sec:parameter_conversion} that 

\begin{equation}
    \left<\frac{\Delta\epsilon}{\epsilon}\right> \approx \frac{(\beta_{\rm u}\gamma_{\rm u})^2 \ln (\bar{\epsilon}_{\rm d} / \bar{\epsilon}_{\rm u})}{\xi},
\label{eq:epsilon_max_rms}
\end{equation}
with $\xi = 55$ is a good approximation across the relevant shock parameter space. Although Equation \eqref{eq:epsilon_max_rms} contains the relativistic four-velocity, it is only valid while $\left< \Delta\epsilon / \epsilon\right> \approx \epsilon_{\rm max}\lesssim 1$.

%%%
%%% COMPARISON TO THERMAL COMPTONIZATION
%%%
\subsection{Modelling an RMS as thermal Comptonization}

The energy gain process described in Section \ref{Sec:energy_gain_RMS} looks strikingly similar to thermal Comptonization on hot electrons \citep[see e.g.,][]{RybickiLightman1979}. Consider a hot cloud of nonrelativistic electrons at a constant temperature $\theta = kT/m_e c^2 \ll 1$ where $k$ is the Boltzmann constant and $T$ is the temperature, with injection of low energy photons ($\epsilon \ll \theta$) into the cloud, and an escape probability that is energy independent. The low-energy photons will gain a relative energy per scattering of $\Delta\epsilon/\epsilon \approx 4 \theta$, and the energy gain continues until balanced by recoil losses at $\epsilon_{\rm max} \approx \Delta\epsilon/\epsilon \approx 4 \theta$. Such Comptonization is described by the Kompaneets equation, with a source term $s$ for the photon injection and escape from the cloud

\begin{equation}
    t_{\rm sc} \left(\frac{\partial n}{\partial t}\right) = \frac{1}{\epsilon^2}\frac{\partial}{\partial \epsilon}\left[\epsilon^4\left(\theta \frac{\partial n}{\partial \epsilon} + n\right)\right] + s,
\label{eq:kompaneets_planar}
\end{equation}

\noindent where $t_{\rm sc} = \lambda/c$ is the Thompson scattering time and $n$ is the photon occupation number. Stimulated scattering ($\propto n^2$) has been omitted in Equation~\eqref{eq:kompaneets_planar}, as this effect is insignificant as long as the occupation number is small, $n \ll 1$, which is true for the non-thermal emission considered here.

\begin{figure}
\begin{centering}
    \includegraphics[width=\columnwidth]{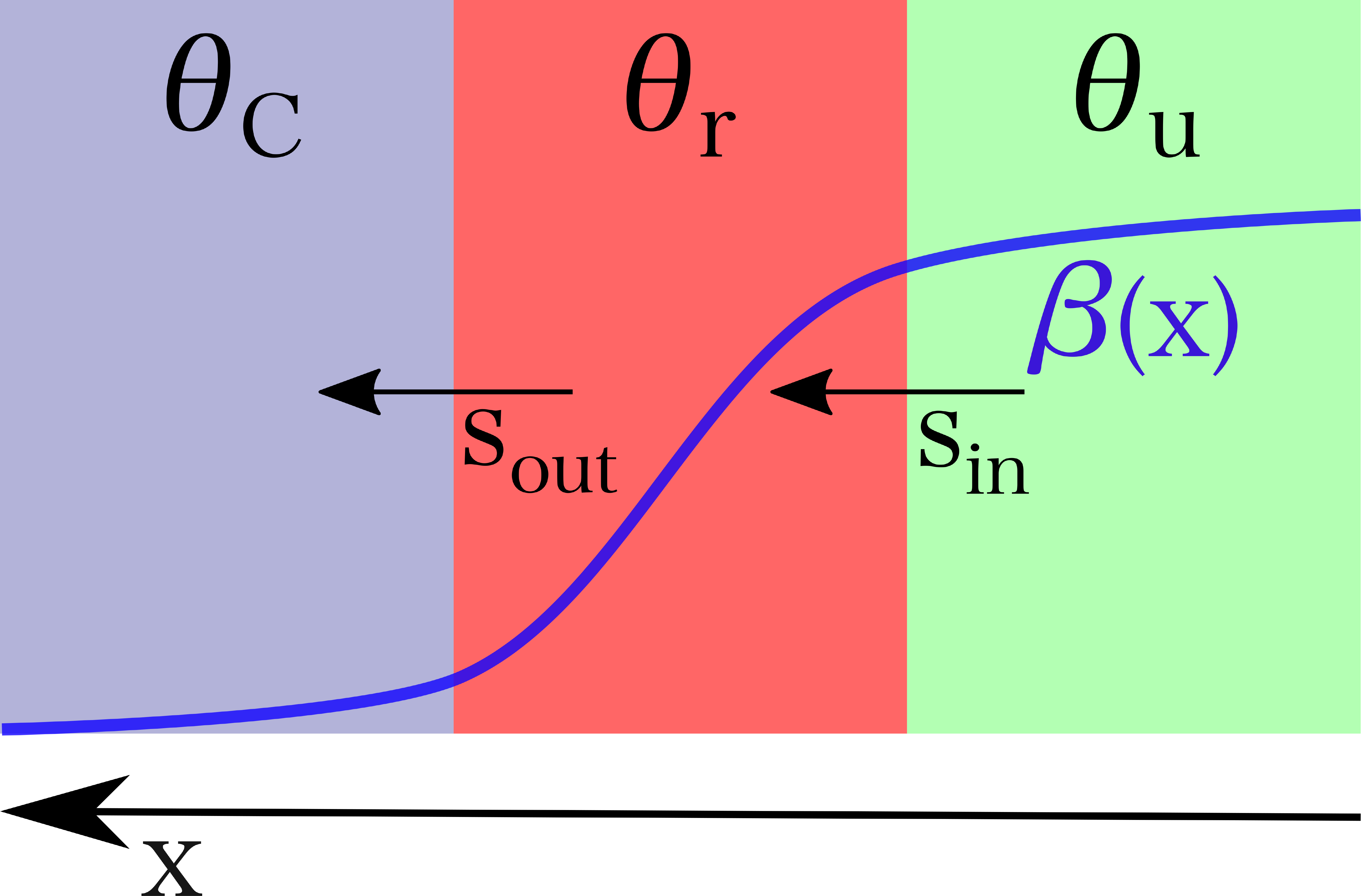}
    \caption{Schematic of the KRA: green indicates the upstream zone, red the RMS zone, and purple the downstream zone. In each zone, photons interact with a population of thermal electrons with an effective temperature, $\theta$. Dissipation occurs in the RMS zone by prescribing the electrons a high temperature $\theta_{\rm r} \gg \theta_{\rm u}$. The zones are connected via source terms, $s$. The overlaid blue line is a rough indication of the velocity profile, $\beta(x)$, across a real RMS, where $x$ is the spatial coordinate.}
    \label{Fig:Threezone}
\end{centering}
\end{figure}

Motivated by the similarities between the two systems, our aim is to construct an approximate RMS model based on the Kompaneets equation. The plasma is split into three discrete zones: the upstream zone, the RMS zone and the downstream zone. The time evolution of the radiation spectrum inside each zone is computed using the Kompaneets equation. Each zone has an effective electron temperature, and the zones are connected via source terms. A schematic of the KRA is shown in Figure~\ref{Fig:Threezone}.

The upstream zone feeds thermal radiation into the RMS zone, which passes radiation on to the downstream zone. Dissipation occurs only inside the RMS zone. This is achieved by prescribing an effective electron temperature $\theta_{\rm r} \approx \frac{1}{4}\Delta \epsilon / \epsilon$, found from Equation \eqref{eq:epsilon_max_rms} as

\begin{equation}
    4\theta_{\rm r} = \frac{(\beta_{\rm u}\gamma_{\rm u})^2 \ln (\bar{\epsilon}_{\rm d} / \bar{\epsilon}_{\rm u})}{\xi}.
\label{eq:equal_energy_gain}
\end{equation}
The subscript r here and henceforth denote quantities in the RMS zone, and the subscripts u and d will be used to denote quantities in the upstream and downstream zones, respectively. Equation \eqref{eq:equal_energy_gain} assures that the maximum photon energy and the energy gain per scattering in the RMS zone mimic those of a real RMS. 
By matching how long photons stay in the shock such that the average downstream energies in the two systems become equal, the evolution of the photon distribution in the KRA will closely match that of a real shock. This is achieved by using appropriate source terms (see Section \ref{sec:source_terms}).
%With the additional condition that the average downstream energies in the two models are equal, the evolution of the photon distribution in both systems are assured to be similar.

The up- and downstream zones do not dissipate energy. Therefore, the temperatures inside these zones equal the radiation Compton temperature $\thetaC$, defined as the electron temperature with which there is no net energy transfer between the photon and electron populations. It is given by

\begin{equation}
    4\theta_\mathrm{C} = \frac{\int \epsilon^4 n \, d\epsilon}{\int \epsilon^3 n \, d\epsilon}.
\end{equation}
where the integrals are taken over all photon energies. The upstream temperature of the radiation\footnote{We use a Wien distribution for the upstream radiation, which is a Planck spectrum with non-zero chemical potential.}, $\theta_{\rm u}$, is a free parameter of the model, but the downstream Compton temperature is not a free parameter, as it is determined by the upstream temperature and the amount of dissipation in the shock. The non-thermal radiation that streams from the RMS zone is accumulated in the downstream zone, where it gradually thermalizes\footnote{That is, high-energy photons preferentially lose energy as they scatter, while low-energy photons gain energy. The net effect is to gradually thermalize the photon distribution, while keeping the average photon energy constant.} via scatterings. The downstream zone contains all photons that passed through the RMS zone, and the degree of thermalization of the radiation inside the downstream zone increases with time. %\deleted{In a real RMS, the Compton temperature may vary across the downstream, specifically if the shock has dissipated a lot of energy. The simplification of discretizing the downstream into a single zone with a common temperature greatly increases the speed of the simulations, but if the shock is very energetic, it may effect the accuracy of the downstream spectrum (see Section \ref{Sec:borderline} for an example).}
%%%
%%%% THE KOMPANEETS SOURCE TERMS
%%%
\subsection{The KRA source terms}\label{sec:source_terms}

The three zones in the KRA are coupled by source terms. Denoting the source of photons that stream into the RMS by $s_{\rm in}$ and the source that streams out of the RMS by $s_{\rm out}$, one gets

\begin{align}
    s_{\rm u} &= - s_{\rm in},\\[2.3mm]
    s_{\rm r} &= s_{\rm in} - s_{\rm out},\label{eq:source_rms}\\[2.3mm]
    s_{\rm d} &= s_{\rm out}.
\end{align}
The probability for a photon to escape the RMS into the downstream is independent of the photon energy. Thus, $s_{\rm out} = k n_{\rm r}$, where $k$ is a constant and $n_{\rm r}$ is the occupation number inside the RMS zone. In this scenario, one can show \citep[e.g.,][]{RybickiLightman1979} that the steady state solution to the Kompaneets equation inside the RMS zone is a power-law distribution, $n \propto \epsilon^{-\alpha}$, with $\alpha = 3/2 \pm (9/4 + k/\theta_{\rm r})^{1/2}$. In analogy with \citet{RybickiLightman1979}, we identify the RMS zone $y$-parameter as $y_{\rm r} = 4\theta_{\rm r} / k$.
%
%\begin{equation}
%    \int_0^\infty \epsilon^2 \left( s_{in} + s_{out} \right) %d\epsilon = 0.
%    \label{eq:rms_photon_conservation}
%\end{equation}
%
\noindent Therefore,

\begin{equation}
    s_{\rm out} = \left(\frac{4\theta_{\rm r}}{y_{\rm r}}\right) n_{\rm r}.
\end{equation}

The RMS zone $y$-parameter determines how much time photons spend inside the shock and it is, therefore, a measure of the average photon energy gain inside the RMS. As such, $y_{\rm r}$ sets the hardness of the non-thermal spectrum that is injected into the downstream. A value of $y_{\rm r} = 1$ corresponds to a flat $\nu F_\nu$-spectrum, with larger values of $y_{\rm r}$ giving harder slopes. The value of $y_{\rm r}$ in the KRA is chosen such that the average downstream photon energy obtained equals that of a real RMS. The full conversion between the parameters that specify the RMS and the corresponding KRA parameters is shown in Appendix \ref{sec:parameter_conversion}.
%Indeed, while Equation \eqref{eq:equal_energy_gain} assures that the energy gain per scattering and the maximum energy obtained in the RMS zone mimic those of a real RMS, the value of $y_{\rm r}$ is chosen such that the number of scatterings in the two systems are similar. It is also a measure of the average energy gain inside the RMS. As such, $y_{\rm r}$ determines the hardness of the non-thermal RMS zone spectrum that is injected into the downstream. A value of $y_{\rm r} = 1$ corresponds to a flat $\nu F_\nu$-spectrum, with larger values of $y_{\rm r}$ corresponding to harder slopes.

%The value of $y_{\rm r}$ is chosen such that the number of scatterings in the RMS zone is similar to that of a real RMS. Together with Equation \eqref{eq:equal_energy_gain}, which assures that the energy gain per scattering and the maximum energy are similar for the two systems, the RMS zone mimics the behaviour of a real RMS.

Requiring that the photon number inside the RMS zone is conserved one finds from Equation \eqref{eq:source_rms}

\begin{equation}
    s_{\rm in} = \left(\frac{4\theta_{\rm r}}{y_{\rm r}}\right) \left(\frac{\int \epsilon^2 n_{\rm r} d\epsilon}{\int \epsilon^2 n_{\rm u} d\epsilon}\right) n_{\rm u},
\end{equation}
where the integrals are again taken over all photon energies.

\subsection{Estimating the KRA upper speed limit}
\label{subsec:estimating_kra_upper_speed_limit}

The energy gain inside the RMS qualitatively changes when the relative energy gain per scattering, $\Delta\epsilon / \epsilon$, becomes comparable to unity. This is because the upper photon energy inside the shock is $\epsilon_\mathrm{max} \approx \Delta\epsilon / \epsilon$, and the radiative transfer is different for such high-energy photons. In particular, Klein-Nishina effects modify the scattering cross-section and $\gamma\gamma$-pair production is triggered. Furthermore, shocks with such a high energy gain per scattering will have a narrow width (comparable to a few photon mean free paths), which makes the radiation anisotropic. The KRA is therefore limited to modelling shocks with $\Delta\epsilon / \epsilon \lesssim 1$. This corresponds to (see Equation \eqref{eq:epsilon_max_rms})

\begin{equation}
    \left(\beta_{\rm u} \gamma_{\rm u}\right)^2 \lesssim \frac{\xi}{\ln\left(\bar{\epsilon}_{\rm d}/\bar{\epsilon}_{\rm u}\right)},
    \label{eq:speed_limit}
\end{equation}

\noindent with $\xi \approx 55$.

The ratio of average downstream to upstream photon energies can vary significantly but enters Equation~\eqref{eq:speed_limit} only as a logarithmic factor. For a typical energy ratio of $\bar{\epsilon}_{\rm d}/\bar{\epsilon}_{\rm u} = 10^2$, one gets an upper velocity limit of $\beta_{\rm u} \gamma_{\rm u} \approx 3.5$. Thus, the KRA is expected to be applicable to shocks with $\beta_{\rm u} \gamma_{\rm u} \lesssim 3$, with the exact value only marginally dependent on the shock parameters.

\subsection{Quasi-thermal RMS spectra}\label{Sec:quasi_thermal}
%\deleted{The radiation inside an RMSs can have a wide variety of spectral shapes, ranging from highly non-thermal to almost thermal.}
The radiation in the downstream of an RMS becomes quasi-thermal if the energy dissipation per photon is either very low or very high. In the former case, the upstream photon distribution is largely unaltered, while in the latter case, the photons gain so much energy that they pile up in a thermal Wien distribution around $\epsilon_{\rm max}$ (i.e., saturated Comptonization). In both cases, the radiation relaxes to a near thermal distribution after a few scatterings in the downstream, at which point the information from the shock is all but lost. When fitting to data, such shocks are almost indistinguishable from each other and from outflows where no dissipation occurred. As such, they are less interesting from an observational perspective.

Shocks with small photon-to-proton ratios, $n_\gamma/n_{\rm p}$, where $n_\gamma$ and $n_{\rm p}$ are the photon and proton number densities, respectively, tend to have more thermal-like spectra inside the RMS. This is because the average downstream photon energy $\bar{\epsilon}_{\rm d}$ is inversely proportional to the photon-to-proton ratio (i.e., more photons sharing the same shock kinetic energy), while the maximum photon energy $\epsilon_{\rm max}$ is proportional to the logarithm of $\bar{\epsilon}_{\rm d}$ (see Equation~\eqref{eq:epsilon_max_rms}). Thus, as the photon number shrinks, $\bar{\epsilon}_{\rm d}$ increases faster than $\epsilon_{\rm max}$, until the spectrum appears quasi-thermal with $\bar{\epsilon}_{\rm d} \sim \epsilon_{\rm max}$.

For $\bar{\epsilon}_\mathrm{d} \gg \bar{\epsilon}_\mathrm{u}$, the average downstream photon energy is

\begin{equation}
    \bar{\epsilon}_{\rm d} \approx \left(\gamma_{\rm u} - 1\right) \frac{m_{\rm p}}{m_{\rm e}} \frac{n_{\rm p}}{n_\gamma},
\end{equation}

\noindent where $m_{\rm p}$ is the proton mass. Equating $\bar{\epsilon}_\mathrm{d}$ to $\epsilon_\mathrm{max} \approx \Delta\epsilon / \epsilon$ and solving for $n_\gamma/n_{\rm p}$, one gets

\begin{equation}
\label{eq:zeta_crit}
    \frac{n_\gamma}{n_{\rm p}} \approx \frac{\gamma_{\rm u} - 1}{(\beta_{\rm u}\gamma_{\rm u})^2} \frac{m_{\rm p}}{m_{\rm e}} \frac{\xi}{\ln\left( \bar{\epsilon}_{\rm d}/\bar{\epsilon}_{\rm u} \right)}.
\end{equation}

\noindent Consider the limit $\beta_{\rm u} \ll 1$, which gives $(\gamma_{\rm u} - 1)/(\beta_{\rm u}\gamma_{\rm u})^2 \approx 1/2$. With $\xi = 55$, and a typical energy ratio of $\bar{\epsilon}_{\rm d}/\bar{\epsilon}_{\rm u} = 10^2$, one finds a critical photon-to-proton ratio of $n_\gamma/n_{\rm p} \approx 1.1 \times 10^4$. Shocks that have photon-to-proton ratios close to this value will result in quasi-thermal radiation spectra. Below, we illustrate this fact with a simulation that has $n_\gamma/n_{\rm p} \approx 4 \times 10^4$.

%\CL{\deleted{From an observational point of view, thermal-looking spectra are less interesting. This is because any information regarding the shock history is largely gone; a large combination of shock parameters can give rise to thermal spectra. As an example, the radiation from a shock with essentially arbitrary parameters will appear thermal given that the shock occurred deep enough inside the GRB jet, as the shocked radiation will thermalize via scatterings before reaching the jet photosphere.}}

\subsection{Comparing the Kompaneets RMS approximation to full RMS simulations}
%%%%
%%%% FIGURES
%%%%
\begin{figure*}
\begin{centering}
    \includegraphics[width=\columnwidth]{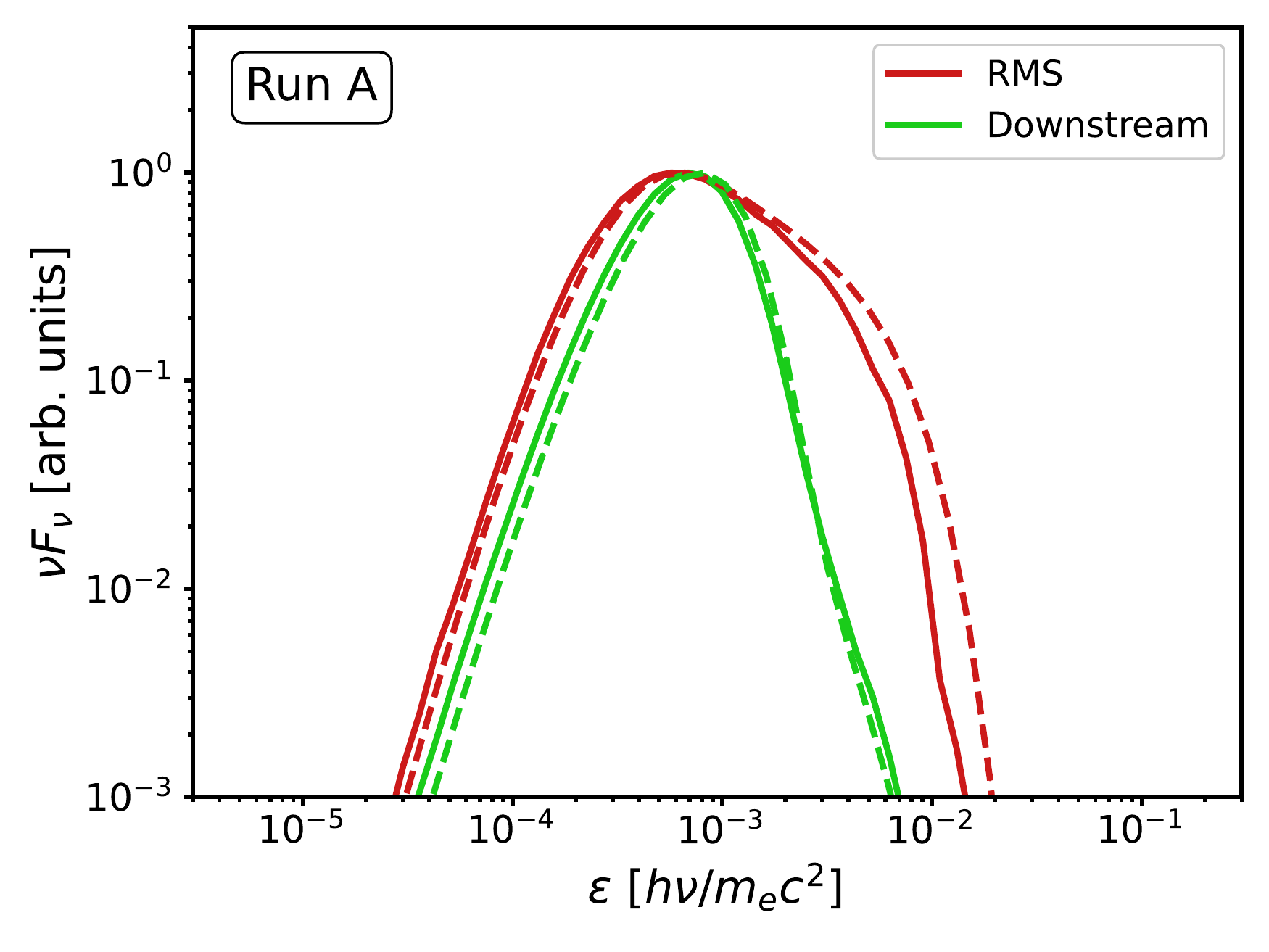}
    \includegraphics[width=\columnwidth]{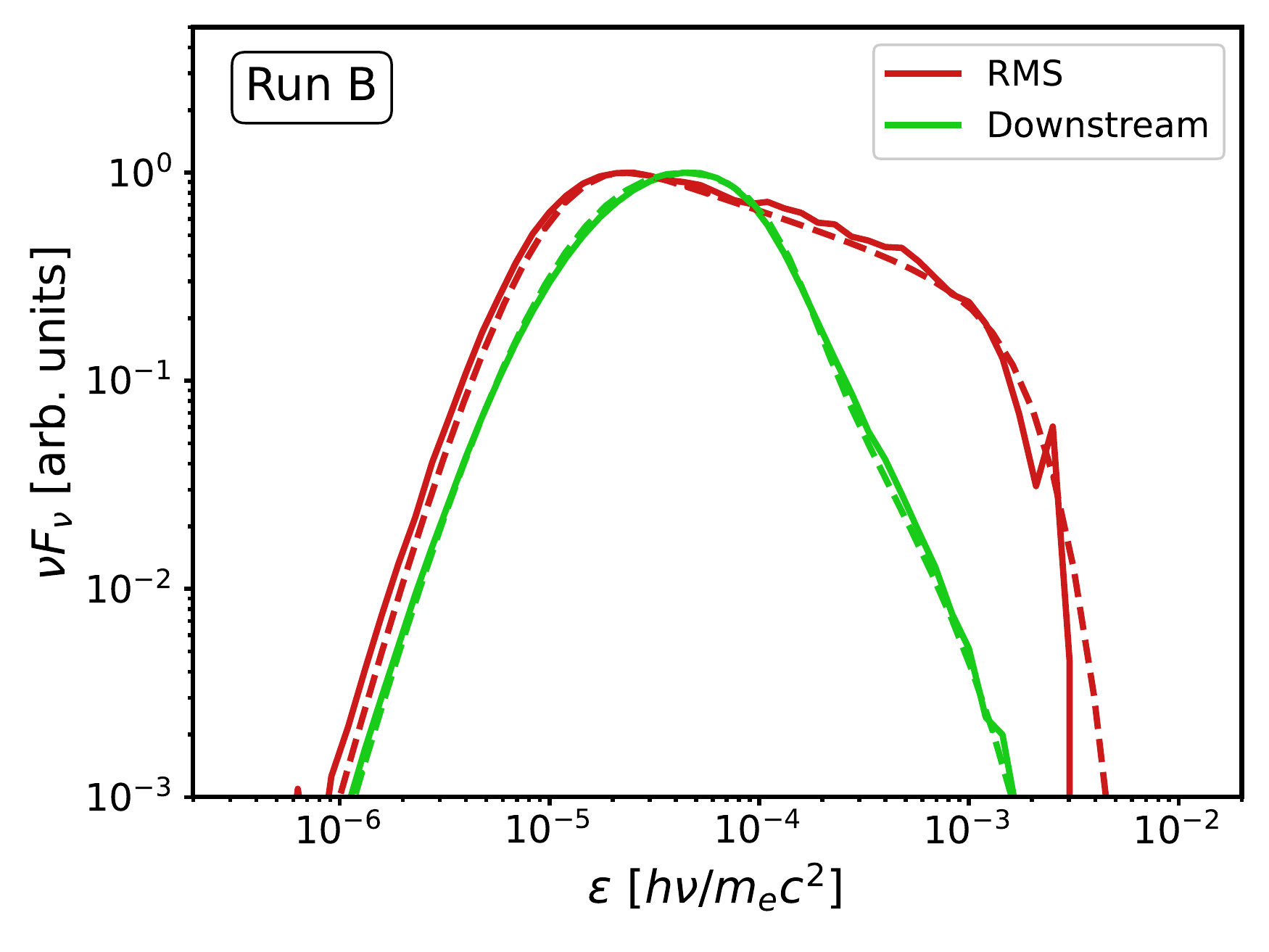}\\
    \includegraphics[width=\columnwidth]{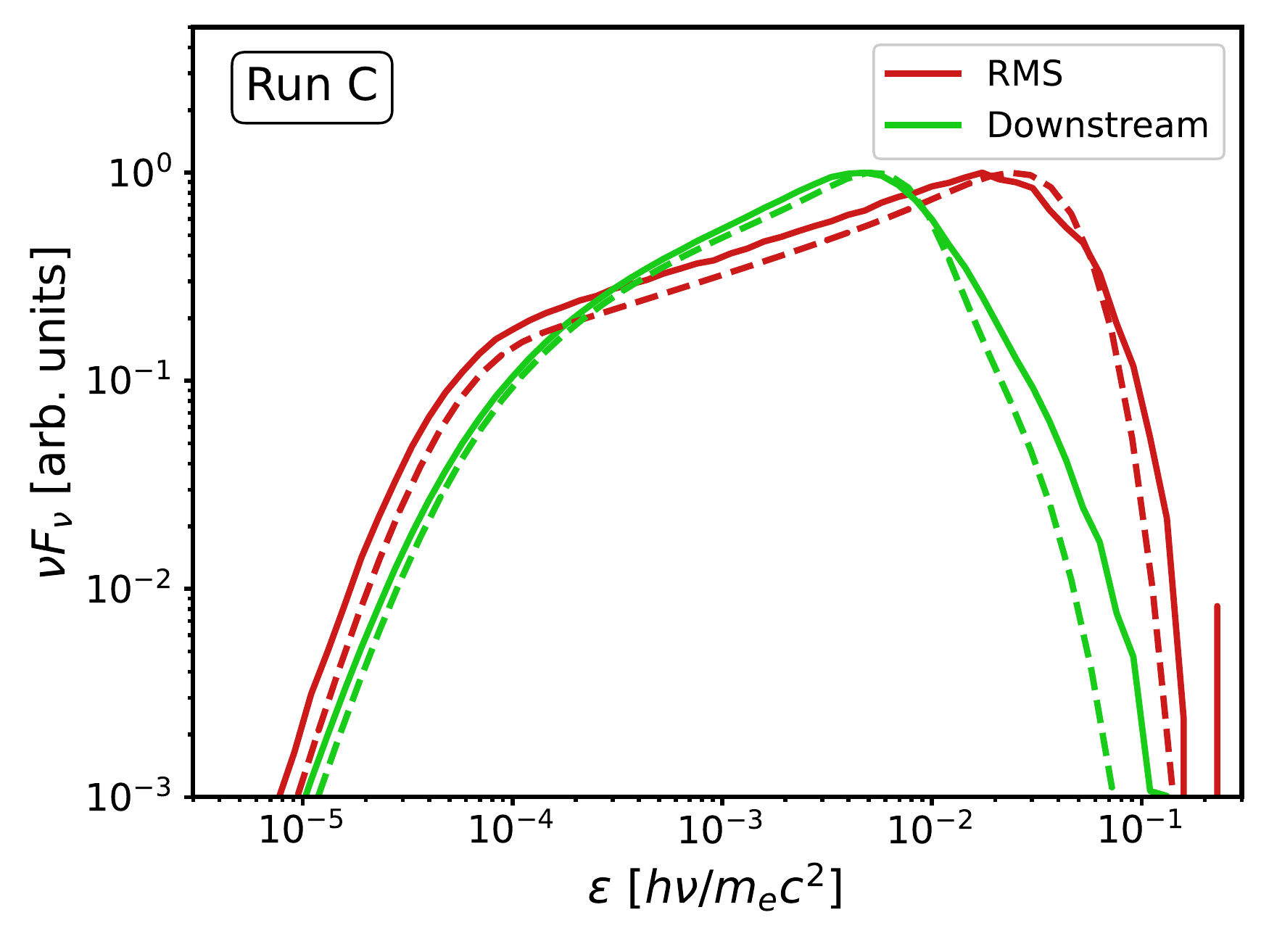}
    \includegraphics[width=\columnwidth]{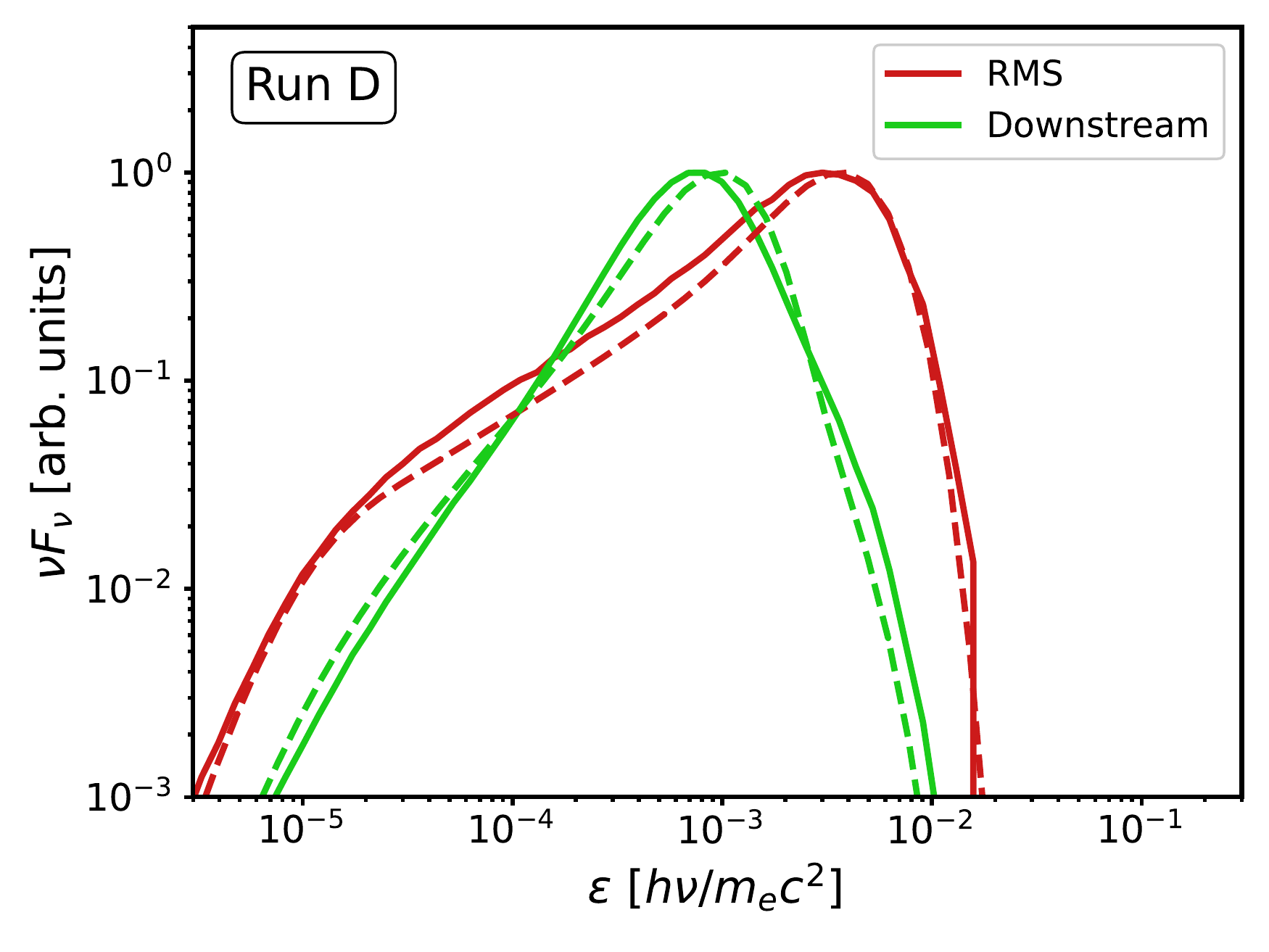}\\
    \includegraphics[width=\columnwidth]{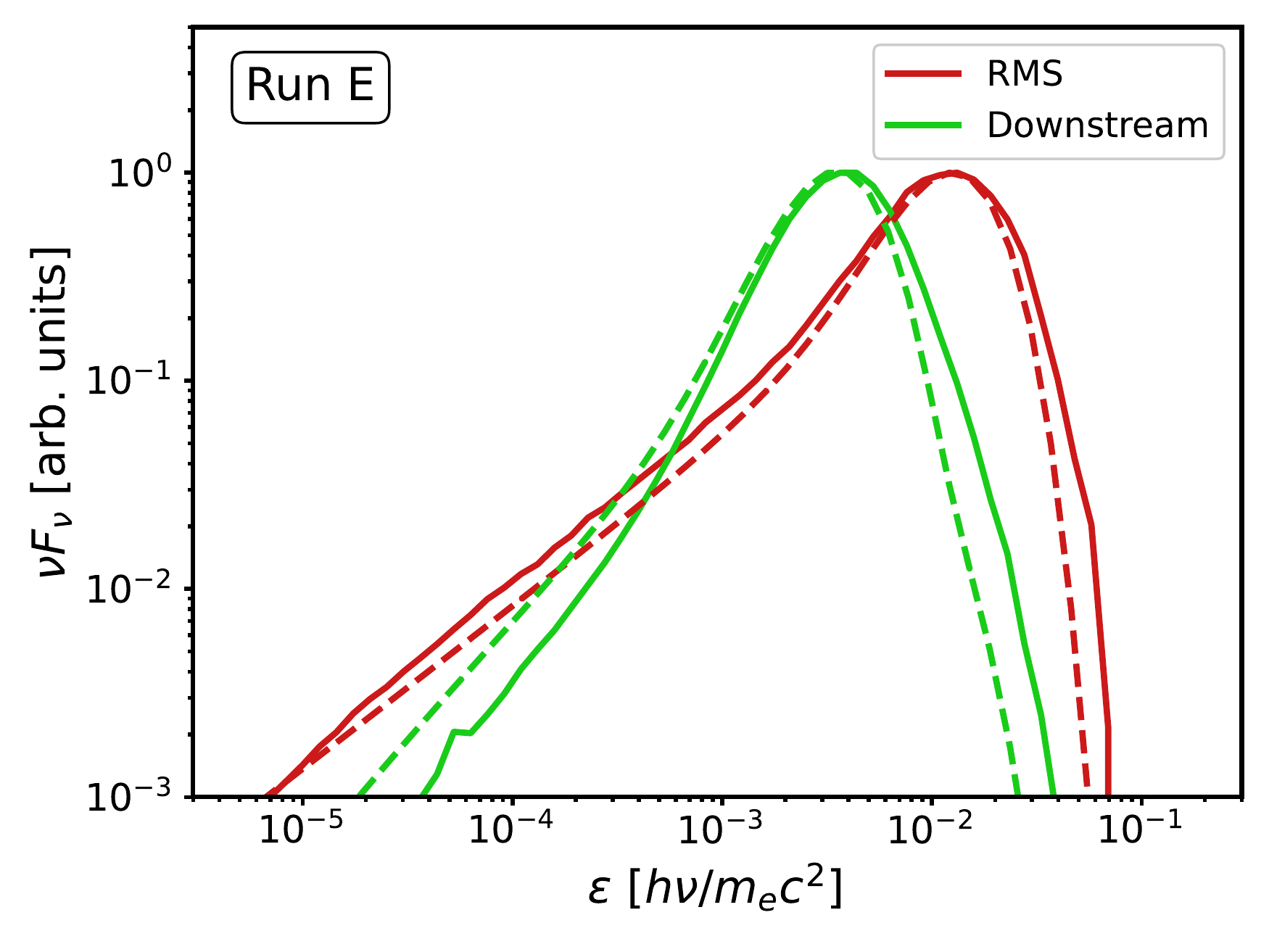}
    \caption{Comparison of RMS and downstream spectra for Runs A--E as indicated in the panels. Solid lines show the spectra from the full radiation hydrodynamics code \radshock\ and dashed lines show spectra from the KRA code \komrad. The parameter values for the runs are given in Tables \ref{tab:komrad_params} and \ref{tab:radshock_params}.}
    \label{Fig:RunAE}
\end{centering}
\end{figure*}
\begin{figure}
\begin{centering}
    \includegraphics[width=\columnwidth]{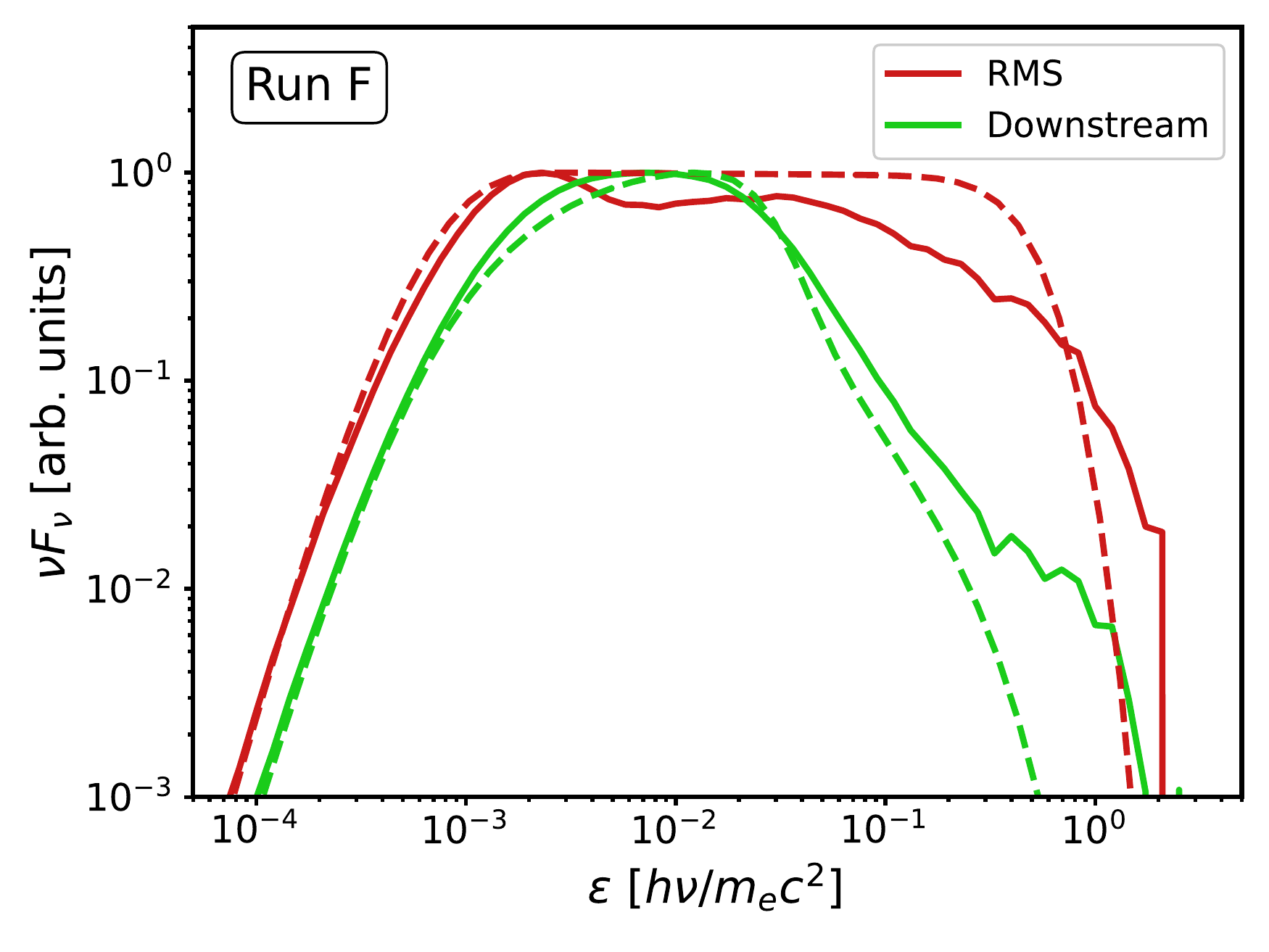}
    \caption{Similar to Figure \ref{Fig:RunAE} but for a mildly relativistic RMS with $\beta_{\rm u}\gamma_{\rm u}=3$ (Run F). The RMS width is only a few optical depths wide and photons can easily diffuse in and out of the shock. Thus, the comparison to a discrete, well defined zone in \komrad\ becomes less accurate. In the downstream, Klein-Nishina effects suppresses the cooling of the highest energy photons. Parameter values for the run are given in Tables \ref{tab:komrad_params} and \ref{tab:radshock_params}.}
    \label{Fig:RunF}
\end{centering}
\end{figure}
%The upper panel of Figure~\ref{fig:comparison_good} shows the comparison of the radiation spectra inside the RMS. 

In this subsection, we will compare the spectrum inside the RMS and downstream regions as computed by the two codes \radshock\ and \komrad, the latter implementing the KRA. The \radshock\ code is a special relativistic, Lagrangian radiation hydrodynamics code \citep{Lundman2018}. The radiation field is computed using the Monte Carlo method, which self-consistently connects to the hydrodynamics via energy and momentum source terms. The RMS is set up by smashing plasma into a wall boundary condition, and allowing the code to relax into an RMS that propagates steadily away from the wall. For the case of a thermal upstream radiation spectrum, the RMS solution is fully specified by three parameters. These can be taken to be the temperature of the upstream radiation, $\theta_{\rm u}$, the speed of the upstream plasma relative to the shock, $\beta_{\rm u}$, and the photon-to-proton ratio, $n_\gamma/n_{\rm p}$, inside the upstream \citep{Lundman2018}.

\komrad\ implements the KRA described in the previous subsections, evolving the radiation in the RMS zone and the downstream zone\footnote{The upstream zone has no need for a Kompaneets solver, as the radiation is assumed to be thermal and the shape of the spectrum is known analytically.} using Kompaneets solvers \citep[e.g.,][]{ChangCooper1970}. 
We choose the following three parameters to describe the RMS in \komrad : the temperature of the upstream photons, $\theta_{\rm u, K}$ (where the subscript K indicates \komrad), the effective electron temperature inside the RMS zone, $\theta_{\rm r}$, and the Compton $y$-parameter of the RMS zone, $y_{\rm r}$. As mentioned above, the conversion from the KRA parameters to the corresponding RMS parameters is given in Appendix~\ref{sec:parameter_conversion}. A non-trivial point is that the two codes will have somewhat different upstream temperatures. This is because plasma compression inside the RMS will increase the internal energy density and shift the upstream spectral peak, and no analogue to this compression exists for the KRA.

A simulation is fully specified by the three shock parameters and the total simulation time $t/t_{\rm sc}$. %, $t_{\rm sc} = \lambda/c = 1/(n_e \sigma_\mathrm{T} c)$).
The simulation time affects the degree to which the downstream has been thermalized. To highlight the similarities between the downstream spectra, we do not include the radiation produced during the initial RMS formation. This is because the formation of the shock is different between the simulations. Therefore, the simulation time starts when the RMS is already in steady state.
%\replaced{if the radiation inside the RMS has had time to reach a steady state, and also how much radiation has been collected in the downstream. Here, we run the codes for long enough to find a steady state spectrum inside the RMS, but not long enough for the downstream spectrum to become fully thermalized (which would make code comparisons uninteresting).}{the degree to which the downstream has been thermalized. To highlight the similarities between the downstream spectra, we do not include the radiation that escapes the shock before the RMS has reached steady state. This is because the formation of the shock is slightly different between the simulations. Therefore, the simulation time starts when the RMS is already in steady state.}

The RMS transition region in \radshock\ is continuous and it is not obvious a priori what part should be compared to the RMS zone in \komrad. For the comparison, we chose the radiation in \radshock\ that is located at the point where the shock has just finished dissipating all incoming energy, as this represents the spectrum that is injected into the downstream. This location is determined as the point where the average photon energy has reached its downstream value. The plasma that has passed through this location after the start of the simulation time belongs to the downstream. At the end of the simulation, the radiation inside the downstream is collected and its spectrum computed.

Figures~\ref{Fig:RunAE} and ~\ref{Fig:RunF} show the results of six different shock simulations, labeled Run A--Run F. The \komrad\ parameters for the six runs are shown in Table~\ref{tab:komrad_params} and the \radshock\ parameters are found in Table~\ref{tab:radshock_params}. The simulation parameters were chosen to test the KRA in different regions of the shock parameter space, resulting in differently shaped radiation spectra inside the shock. The \komrad\ parameters for the six runs are calculated from the corresponding \radshock\ parameters using the method described in Appendix \ref{sec:parameter_conversion}. The only free parameter in the conversion is $\xi$ from Equation \eqref{eq:epsilon_max_rms}. All \komrad\ runs are made with $\xi = 55$, as we empirically found this value gave good agreement across the parameter space.

In Figure~\ref{Fig:RunAE}, Run A--Run E are shown in five different panels. The spectra produced by the two codes are remarkably similar, highlighting the close analogy between bulk and thermal Comptonization. We conclude that the KRA can accurately capture the RMS radiation physics.

In Figure~\ref{Fig:RunF}, we show the spectra for Run E, which is a mildly relativistic shock with upstream speed $\beta_{\rm u} \gamma_{\rm u} = 3$ in the shock rest frame. The KRA neglects relativistic effects such as Klein-Nishina suppression and pair production. Furthermore, as shown in e.g., \citet{Ito2018}, anisotropy starts to become important when the shock becomes relativistic. However, \komrad\ can still capture the behavior of mildly relativistic shocks, as long as the photon energies inside the RMS do not exceed the electron rest mass, i.e., as long as $\epsilon_{\rm max} \lesssim 1$ as discussed in Section \ref{subsec:estimating_kra_upper_speed_limit}. %In Figure \ref{fig:comparison_fast_shock}, we show a comparison between the two models when $\Delta \epsilon / \epsilon \sim 1$. As evident from the Figure, this is the point where the approximation starts to deviate from the true solution. 

In this run, the relative energy gain per scattering is close to unity and the shock is only a few Thomson optical depths wide in the \radshock\ simulation. Hence, photons have a high probability of diffusing in and out of the different regions and there are no sharp ``zone boundaries'' in \radshock. Therefore, the comparison to a discrete RMS zone in \komrad\ is less accurate. Furthermore, Klein-Nishina suppression starts to become important. This can be seen in the high-energy tail of the photon distribution in the downstream. The high-energy photons in \radshock\ have cooled less than those in \komrad, due to their lower scattering cross section. However, this effect will likely be unimportant when the KRA is fitted to actual data, as radiation with $\epsilon \lesssim 1$ have time to downscatter to lower energies before reaching the photosphere, even with Klein-Nishina suppression. Furthermore, the high-energy part of the spectrum is often given little weight in the fitting process due to the lower photon counts at high energies in the GBM energy channels \citep{Yu2019}. Overall the approximation is surprisingly accurate even in this case when $\beta_{\rm u}\gamma_{\rm u} = 3$, especially in the downstream zone, which contains the radiation that will later be observed. This indicates that any anisotropy of the radiation field within the shock does not have a major impact on the shape of the spectrum in this case. We conclude that the limit of the KRA is when $\epsilon_{\rm max} \approx \Delta \epsilon / \epsilon \approx 4\theta_{\rm r}$ starts to approach unity. 

%\added{In Figure \ref{Fig:RunF}, we show a run where the photon to proton ratio is close to the critical value given in Equation \eqref{eq:zeta_crit}. Shocks with few photons per proton result in very hard RMS spectra and quasi-thermal downstream spectra. In \komrad, this corresponds to shocks with high $y_{\rm r}$. As discussed in Section \ref{Sec:quasi_thermal}, such spectra are less interesting from an observational point of view as they are hard to distinguish from other thermal and quasi-thermal spectra. Still, \komrad\ still accurately and \radshock\ is good.}
%
\begin{table}[]
    \centering
    \caption{\komrad\ simulation parameters}
    \begin{tabular}{ccccc}
    \hline
         Run & $t/t_{\rm sc}$ & $\theta_{\rm u, K}$ & $R \equiv \theta_{\rm r}/\theta_{\rm u, K}$ & $y_{\rm r}$ \\
    \hline
         A & $5 \times 10^{3}$ & $1.05 \times 10^{-4}$ & 15.3 & 0.56 \\
         B & $1.5 \times 10^{4}$ & $3.35 \times 10^{-6}$ & 110 & 0.70 \\
         C & 320 & $1.73 \times 10^{-5}$ & 522 & 1.58 \\
         D & $5 \times 10^{3}$ & $3.35 \times 10^{-6}$ & 325 & 2.97 \\
         E & 2834 & $6.04 \times 10^{-7}$ & 5644 & 5.6 \\         
         F & 80 & $2.51 \times 10^{-4}$ & 403 & 0.99 \\
    \hline
    \end{tabular}
    \label{tab:komrad_params}
\end{table}
\begin{table}[]
    \centering
    \caption{\radshock\ simulation parameters}
    \begin{tabular}{ccccc}
    \hline
         Run & $t/t_{\rm sc}$ & $\theta_{\rm u}$ & $\beta_{\rm u}$ & $n_\gamma/n_{\rm p}$ \\
    \hline
         A & $5 \times 10^{3}$ & $6.13 \times 10^{-5}$ & 0.490 & $5.47 \times 10^5$ \\
         B & $1.5 \times 10^{4}$ & $1.89 \times 10^{-6}$ & 0.224 & $1.70 \times 10^6$ \\
         C & 320 & $8.86 \times 10^{-6}$ & 0.610 & $4.82 \times 10^{5}$ \\
         D & $5 \times 10^{3}$ & $1.75 \times 10^{-6}$ & 0.228 & $9.00 \times 10^4$ \\
         E & 2834 & $3.14 \times 10^{-7}$ & 0.303 & $4.12 \times 10^{4}$ \\
         F & 80 & $1.1 \times 10^{-4}$ & 0.949 & $10^{6}$ \\
    \hline
    \end{tabular}
    \label{tab:radshock_params}
\end{table}
\section{Applying the Kompaneets RMS approximation to a GRB jet}\label{Sec:jet_geometry}
RMSs come with a variety of dynamical behaviors. Explosions, such as supernovae, generate an outward going shockwave. The shockwave propagates through the star until it either breaks out of the stellar surface (i.e., the photosphere), or dissolves as the downstream pressure becomes too small, due to the limited explosion energy budget. A different dynamical behavior is seen in recollimation shocks, which arise as the jet propagates in a confining medium. Such shocks can be approximately stationary with respect to the star and might therefore never break out. However, the radiation that is advected through the recollimation shock is energized, and the emission released at the photosphere can be non-thermal. Yet another behavior is seen in shocks that arise due to internal collisions of plasma inside the jet. When two plasma blobs collide, the plasma in between the blobs is compressed, increasing the pressure adiabatically until the pressure profile is steep enough to launch two shocks. The shocks propagate in opposite directions into the two colliding blobs while sharing a causally connected downstream region. Such shocks cease when they have dissipated most of the available kinetic energy in the two blobs. The time it takes the shocks to cross the blobs is roughly a dynamical time (as they cross the causally connected jet ejecta).

Dynamical effects on the shock structure are important if the shock reaches the jet photosphere. For instance, part of the RMS can transform into a pair of collisionless shocks at the point of breakout when the photons mediating the shock start to leak out toward infinity \citep{LundmanBeloborodov2021}. The KRA is not able to handle such dynamical effects. On the other hand, the KRA is well suited for simulating plasma that is shocked while still being optically thick, and then advected toward the jet photosphere where the emission is later released.

\subsection{A minimal sub-photospheric shock model}

One shock scenario that can be modeled by the KRA is the collision inside the jet of two blobs of similar mass and density, but different speeds. This is a minimal scenario with few free parameters. More complex models with additional parameters can be considered in the future, if the current model fails to fit prompt GRB data. In Appendix~\ref{sec:are_is_relativistic} we compute the speed of the two shocks, the energy dissipated, and the radius at which the shocks have finished dissipating most of the available energy. For blobs of similar properties (i.e., similar mass and density), we show that also the properties of the shocks are similar. In that case, only one of the two shocks have to be explicitly simulated, and the number of model parameters are kept at a minimum.

The KRA is valid for shocks where $\Delta\epsilon/\epsilon \lesssim 1$, as discussed in Section \ref{subsec:estimating_kra_upper_speed_limit}. In Appendix~\ref{sec:are_is_relativistic} we translate this limit to jet quantities in the context of internal shocks. It corresponds to a Lorentz factor ratio of $\Gamma_2/\Gamma_1 \lesssim 30$, where $\Gamma_2$ and $\Gamma_1$ are the lab frame Lorentz factors of the fast and slow blobs, respectively. As an example, $\Gamma_1 = 50$ and $\Gamma_2 = 1000$ produces shocks that the KRA can accurately model.

\begin{figure}
\begin{centering}
    \includegraphics[width=\columnwidth]{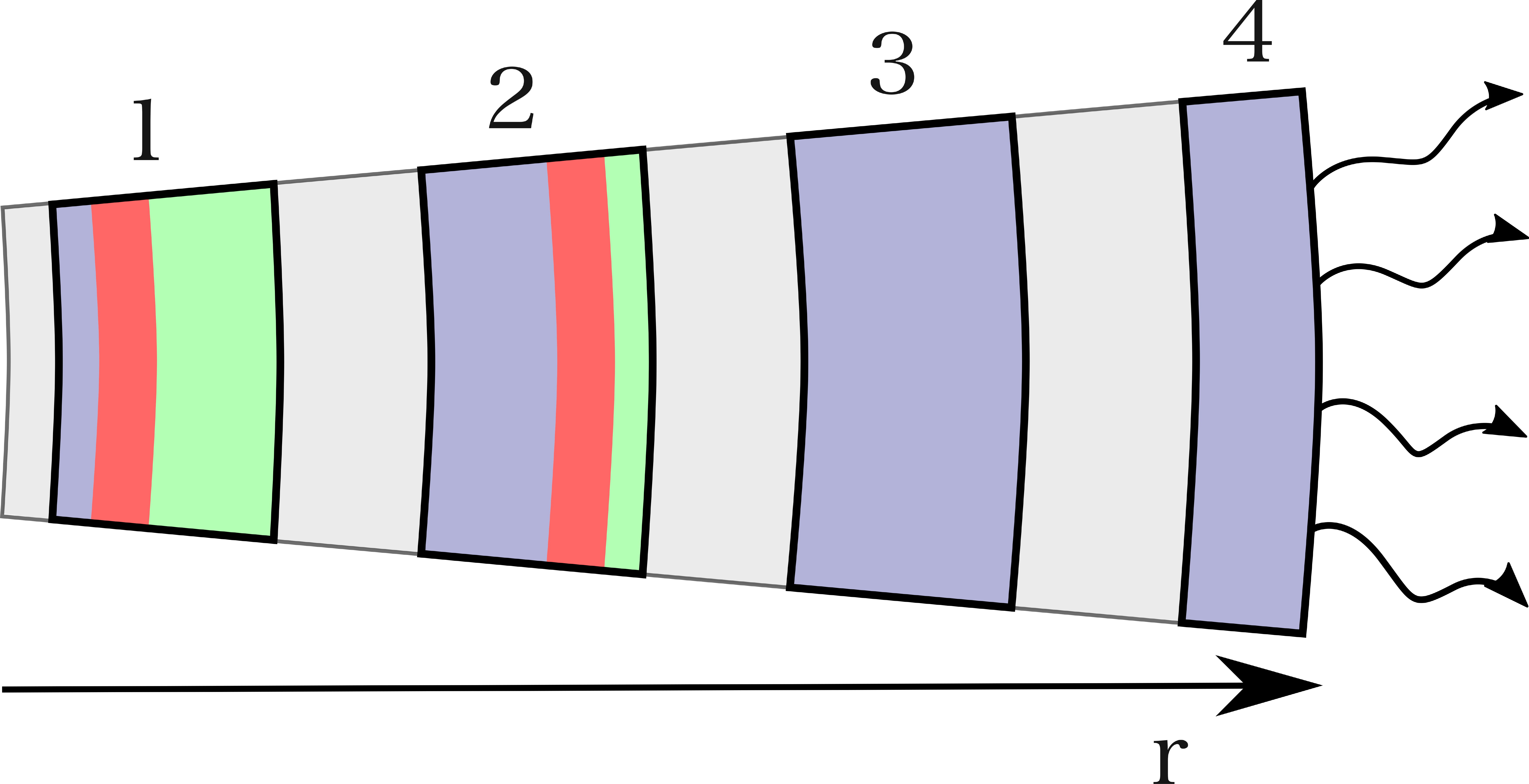}
    \caption{Schematic showing four stages in the evolution of the minimal shock model from left to right. Green, red, and purple indicate the upstream, RMS, and downstream zones, respectively. At some optical depth $\tau_{\rm i}$ (small $r$), the shock is initiated. Photons start to diffuse through the shock region and gain energy. The photons that pass through the RMS are collected in the downstream region, which gain more and more photons (stages 1 and 2). When the RMS has crossed the upstream it dissolves, leaving only the downstream (stage 3). The photons in the downstream continues to scatter until they are released at the photosphere (stage 4).}
    \label{Fig:jet}
\end{centering}
\end{figure}

A schematic illustration of the minimal shock model at four different stages of its evolution is shown in Figure~\ref{Fig:jet}. In the first stage, the two blobs have recently collided. The RMS has started to propagate into the upstream and a few photons have had time to diffuse into the downstream. In the second stage, the shock has almost crossed the upstream. The shock has finished crossing the upstream and dissolved in the third stage, with all photons accumulated in the downstream, and the fourth stage shows the radiation being released at the photosphere. Each of the three zones account for thermalization of the photon spectrum via scatterings and adiabatic cooling.

The time over which shocks dissipate their energy is not a free parameter; it should be found self-consistently from, e.g., hydrodynamical simulations. However, the shock crossing time is always comparable to the dynamical time scale of the jet, which corresponds to a doubling of the jet radius. Therefore, we let the KRA dissipate energy over a doubling of the jet radius in our shock model. This also corresponds to a halving of the optical depth as we consider the case of a conical outflow. After the shock stops dissipating, the downstream plasma containing the shocked radiation is advected toward the photosphere, while it gradually thermalizes through scatterings and cools adiabatically. The simulation ends when the shocked radiation reaches the photosphere.

As mentioned above, we omit photon production by the plasma (i.e., the shock is photon rich). This is a valid assumption, as the advected flux of upstream photons already existing inside the GRB jet is much larger than the number of photons produced by bremsstrahlung or double Compton scattering \citep[e.g.,][]{Bromberg2011b, LundmanBeloborodov2019}. Photon production will occur in the downstream but the time scale for such production is long; the photon spectrum thermalizes into a Wien spectrum via scatterings long before photon production acts to modify the Wien spectrum into a Planck spectrum. As shown by \citet{Levinson2012}, photon production has time to modify the spectrum if the shock occurred at optical depths of $\sim 10^5$. In that case, the radiation will have lost essentially all its energy to adiabatic expansion before reaching the photosphere, and is, therefore, of little interest.

\subsection{KRA implementation in spherical geometry}
A conical jet appears locally as spherically symmetric. The Kompaneets equation inside a steady state, spherical relativistic outflow (with outflow bulk velocity $\beta \rightarrow 1$) is given by Equation (3) of \citet{VurmBeloborodov2016}. With the assumptions of a constant bulk Lorentz factor $\Gamma$, no induced scattering ($n \ll 1$), and no emission or absorption of photons, the Kompaneets equation can be written as

\begin{multline}
	\frac{\partial}{\partial \rbar}\left(\rbar^2 n\right) = \\[2.3mm]
	\frac{1}{\epsilon^2}\frac{\partial}{\partial \epsilon}\left\{\frac{\epsilon^4}{\rbar^2}\left[\theta \frac{\partial \left(\rbar^2 n\right)}{\partial \epsilon} + \left(\rbar^2 n\right)\right] + \frac{2}{3} \frac{\epsilon^3\left(\rbar^2 n\right)}{\rbar}\right\} + s.
\label{eq:kompaneets_spherical}
\end{multline}

\noindent where $\bar{r} = r/R_{\rm ph}$ is a normalized radius and $R_{\rm ph}$ is the radius of the photosphere. The normalized radius equals $\bar{r} = 1/\tau$, where $\tau = \sigma_\mathrm{T} n_{\rm e} r / \Gamma$ is the optical depth of the jet (not to be confused with the optical depth of the RMS) with $\sigma_{\rm T}$ being the Thomson cross section and $n_{\rm e}$ the electron number density. The comoving time coordinate in the Kompaneets equation has here been re-written into a lab frame radial coordinate (using $t = r/\Gamma c$).

The last term in the curly brackets of Equation \eqref{eq:kompaneets_spherical} accounts for adiabatic cooling of the spectrum.\footnote{The Kompaneets solver method described in \citet{ChangCooper1970} with a small grid size is not directly applicable here anymore, since no stationary solution to the Kompaneets equation exists when adiabatic cooling is included. However, increasing the energy grid size assures that convergence is obtained.} In the optically thick regime, this causes the average energy of the photon distribution to decrease as $\bar{r}^{-2/3}$, while the shape of the spectrum is preserved. When the photons start to decouple close to the photosphere $(\tau \gtrsim 1)$, the evolution changes and the idealized cooling of $\bar{r}^{-2/3}$ is no longer valid \citep{Peer2008, Beloborodov2011}. To account for this, we numerically stop the cooling at an optical depth of $\tau = 3$. The total adiabatic cooling of the photon distribution is then similar to that of a real spectrum where the proper radiation transfer is taken into account \citep[see,][]{Beloborodov2011}. Scattering is incorporated until $\tau = 1$.

\subsection{Lab frame transformation of the simulated radiation spectrum}
 \label{sec:broad}

The simulation outputs the comoving radiation spectrum at the jet photosphere. The simplest approximate transformation to the lab frame involves multiplying all photon energies by a factor $\Gamma$ (the Doppler boost for a typical photon), where $\Gamma$ is the Lorentz factor of the downstream zone. Since $\Gamma$ does not explicitly enter the simulation, it is effectively a post-processing parameter.

In reality, the radiation spectrum broadens somewhat as it decouples from the plasma at the jet photosphere \citep{Peer2008, Beloborodov2010, Lundman2013}. This is because individual photons decouple at different angles to the line-of-sight, which affects their Doppler boosts, and also at different radii, which affects their energy losses due to adiabatic expansion. These effects are important to take into account when performing spectral fits to data, specifically to narrow bursts \citep{Ryde2017}. This spectral broadening can be approximately computed in a post-processing step, under the assumption that the jet Lorentz factor $\Gamma$ is constant at the photosphere. The post-processing calculation is fairly long, and will be described in full detail elsewhere.

The Kompaneets equation without the induced scattering term is linear in the photon occupation number $n$. Therefore, the total photon number of the simulation is also a free parameter, which effectively makes the normalization of the GRB luminosity a post-processing parameter.

\subsection{New parameters based on parameter degeneracy}
In the case of planar geometry described in Section \ref{sec:KRA}, the KRA has three parameters: $\theta_{\rm u}$, $\theta_{\rm r}$, and $y_{\rm r}$. In the case of a jet, the optical depth of the jet where the shock is initiated, $\tau_{\rm i}$, is an additional parameter. Together, they determine the shape of the photospheric spectrum in the rest frame of the outflow. Additionally, as described in the subsection above, the bulk Lorentz factor $\Gamma$ and the luminosity $L_\gamma$ are two post-processing parameters that shifts the observed spectrum in energy and flux.

However, there exists a degeneracy in the current parameter set. As long as the product $\tau_{\rm i}\theta_{\rm r}$, the ratio $\theta_{\rm r}/\theta_{\rm u}$, and $y_{\rm r}$ remain unchanged, the spectral \textit{shape} in the rest frame of the outflow is identical. With the translational freedoms given by $\Gamma$ and $L_\gamma$, this becomes degenerate. Therefore, a more suitable set of parameters are $\tau\theta \equiv \tau_{\rm i}\theta_{\rm r}$, $R \equiv \theta_{\rm r}/\theta_{\rm u}$, and $y_{\rm r}$. This brings the number of simulation parameters down from four to three, which simplifies the process of table model building.

The degeneracy can be understood as follows. Imagine $\tau_{\rm i}$ is increased by some factor $f$ but $\theta_{\rm r}$ and $\theta_{\rm u}$ are decreased by the same amount. The evolution of the photon distribution with optical depth is slower, as the energy transfer per scattering is proportional to the electron temperature (when the photon gains energy in the scattering) or the photon energy (when the photon loses energy), both of which have decreased by a factor $f$. However, the number of scatterings is $f$ times larger, so the relative energy transfer is the same, i.e., the spectrum evolves similarly but is a factor $f$ lower in energy. The net effect is that the evolution of the whole system is equivalent. The shape of the photospheric spectrum is identical, but shifted down in energy by a factor $f^{5/3}$, where an additional factor $f^{2/3}$ comes from the increased adiabatic cooling. The degeneracy in number of scatterings and energy gain per scattering is not unique to our model. Indeed, it is inherent to all jetted RMS models.

\begin{figure}
\begin{centering}
    \includegraphics[width=0.95\columnwidth]{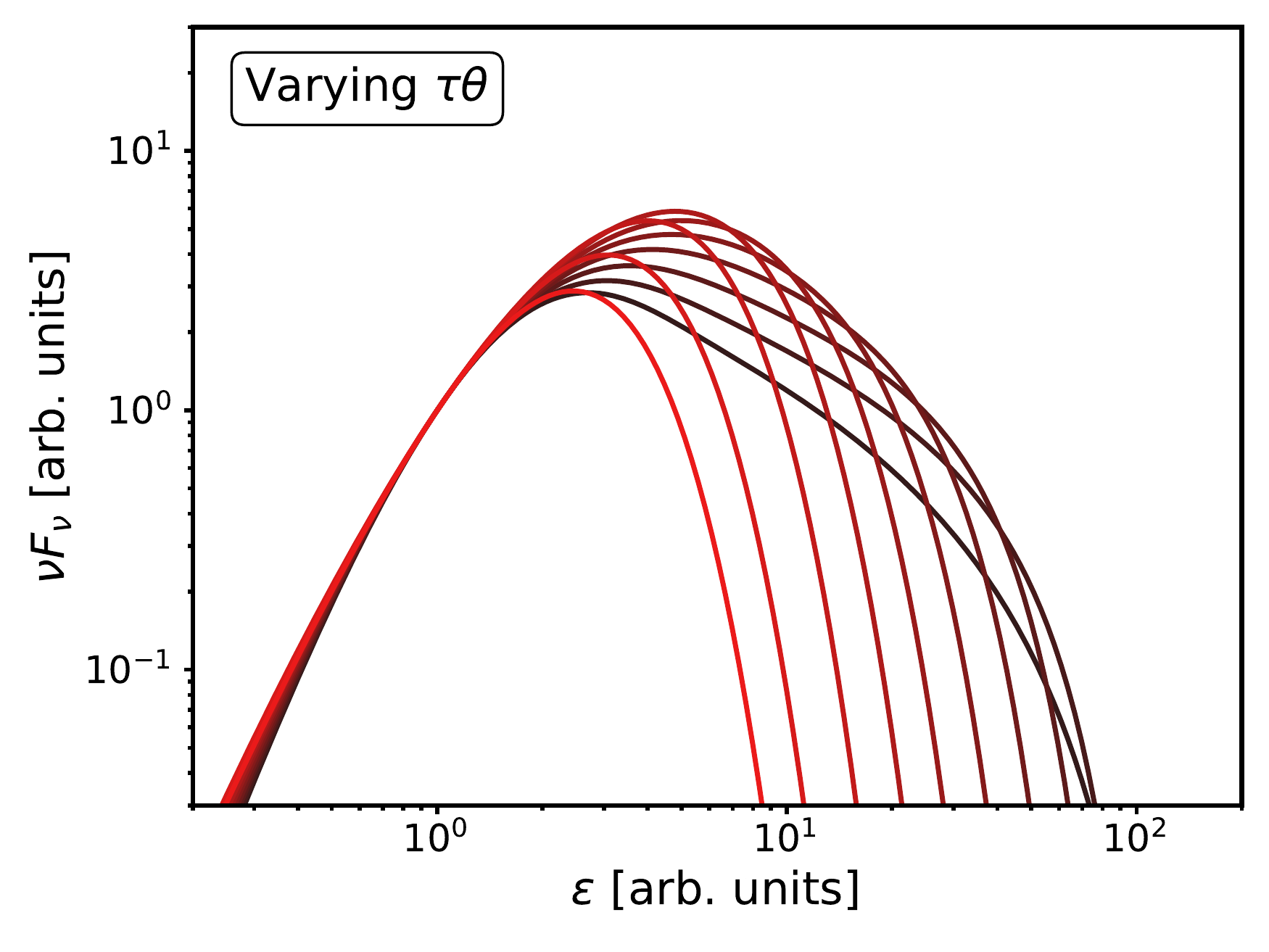}\\
    \includegraphics[width=0.95\columnwidth]{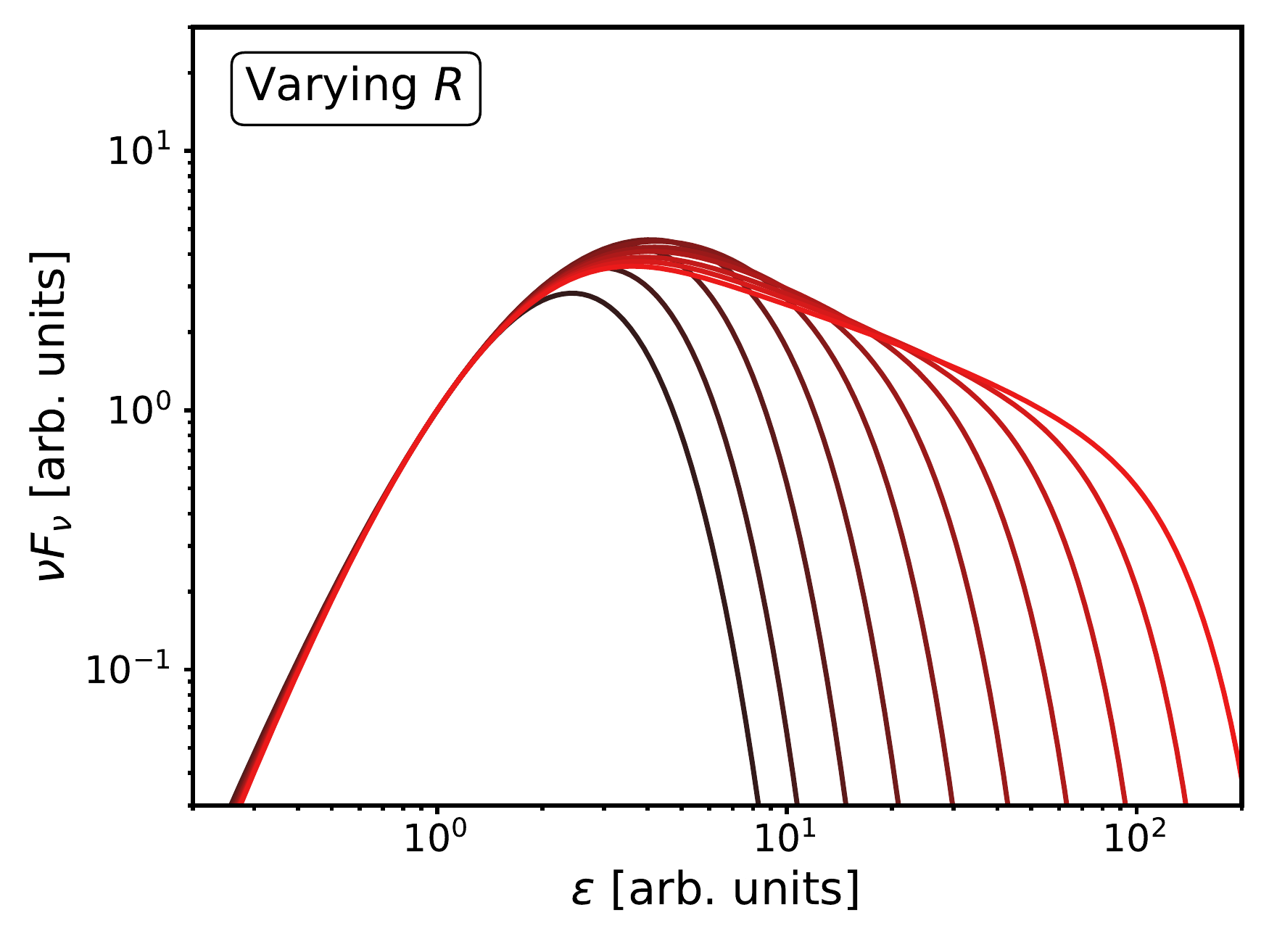}\\
    \includegraphics[width=0.95\columnwidth]{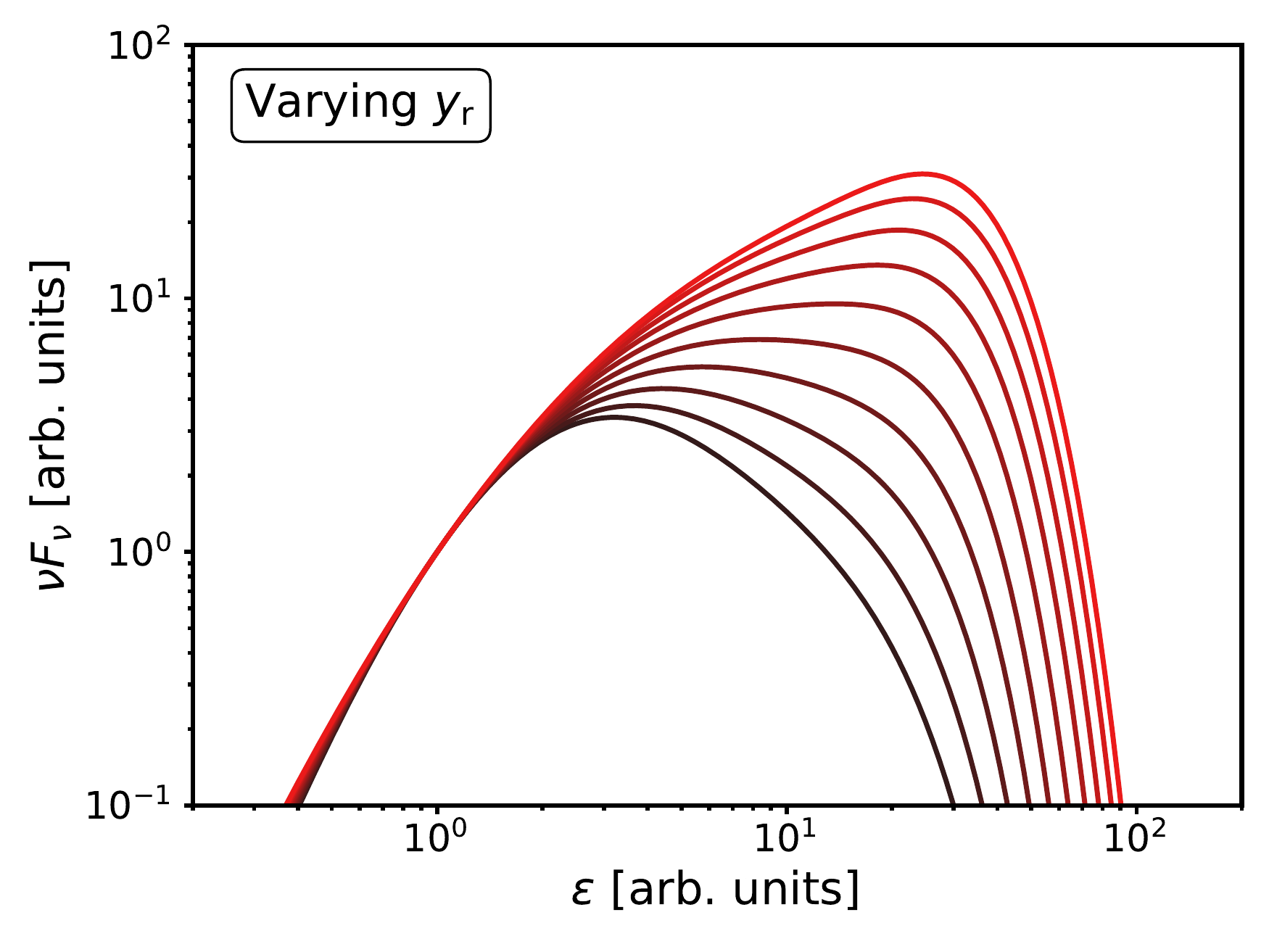}
    \caption{Photospheric spectra generated by \komrad\ in the minimal shock model. In each panel, one of the three parameters $\tau\theta$, $R$, and $y_{\rm r}$, is varied as indicated in the panels, while the other two parameters are kept constant. The constant parameter values are $\tau \theta = 5$, $R = 100$, and $y_{\rm r} = 0.7$. The value of the varying parameter increases from black to red, being evenly log-spaced from $1.5$ to $50$ for $\tau\theta$, $10$ to $10^3$ for $R$, and $0.5$ to $3$ for $y_{\rm r}$. Increasing $\tau\theta$ increases the thermalization. The ratio $R$ determines the separation between the lower and upper cutoff in the spectrum and $y_{\rm r}$ determines the slope of the power-law segment. All spectra have been normalized to unity in photon number ($N_\nu = 1$) at $\epsilon = 1$.}
    \label{Fig:Parameter_interplay}
\end{centering}
\end{figure}

To see how each parameter influences the shape of the released photospheric spectrum, in Figure \ref{Fig:Parameter_interplay} we vary $\tau\theta$ (top panel), $R$ (middle panel), and $y_{\rm r}$ (bottom panel), while keeping the other parameters constant. The value of the parameter being varied increases from black to red. As can be seen in the figure, the combined parameter $\tau\theta$ determines the amount of thermalization after the shock has finished dissipating its energy. A higher $\tau\theta$ implies a higher number of scatterings and/or higher energy transfer per scattering, leading to a faster thermalization. For large $\tau\theta$ the downstream spectrum relaxes to a Wien spectrum, in which case the original shock parameters cannot be retrieved. The ratio $R \equiv \theta_{\rm r}/\theta_{\rm u}$ determines the separation between the lower and upper cutoff in the spectrum. A large $R$ leads to a long power-law segment in the downstream. The slope of the power-law depends on the Compton $y$-parameter $y_{\rm r}$, which is a measure of how much energy is dissipated in the shock. Higher values of $y_{\rm r}$ lead to harder spectra, with $y_{\rm r} = 1$ corresponding to a flat $\nu F_\nu$-spectrum. The spectral broadening discussed in Section \ref{sec:broad} has been omitted in Figure \ref{Fig:Parameter_interplay}, so that the effect of each parameter on the final spectrum is more clearly seen. 

\section{Fitting GRB data with the Kompaneets RMS Approximation}\label{Sec:Example_fit}

As a proof of concept of fitting an RMS model to prompt GRB emission data, we present an analysis of a time-resolved spectrum in GRB 150314A. This luminous burst was observed by the {\it Fermi Gamma-ray Space Telescope} and its Gamma-ray burst monitor (GBM), which covers the energy range of 8 keV--40 MeV. 

GRB 150314A is an example of a GRB pulse in which the spectrum becomes very narrow during a portion of its duration \citep{Yu2019}.
The low-energy photon-index, $\alpha$, of the Band-function \citep{Band1993} reaches a very large value, $\alpha_{\rm max} = -0.27$. Such a large $\alpha$-value strongly suggest a photospheric origin of the emission during the analysed time-bin \citep{Acuner2020}. It is further natural to assume that the same emission mechanism operates throughout a coherent pulse structure such as the one in GRB 150314A \citep{Yu2019}. Therefore, the whole pulse can be argued to be photospheric, even though most of the other time bins have non-thermal spectra ($\alpha \sim -1$). As described in the introduction, these non-thermal spectra must then have been formed by subphotospheric dissipation \citep{ReesMeszaros2005, Ryde2011} with RMSs as the most probable source of dissipation \citep[e.g.,][]{LevinsonBromberg2008, LundmanBeloborodov2019}.

In order to perform fast and efficient fits with the RMS model, we generate synthetic photospheric spectra over a large parameter space using \komrad\ within the minimal shock model as explained above. With the photospheric spectra, we construct a table model in the Multi-Mission Maximum Likelihood Framework \citep[3ML; ][]{Vianello2015}. Here, we include the broadening effect described in Section \ref{sec:broad} by post processing of the spectra. Our initial table model consists of 125 spectra.

%In this Section, we show that the model can be used to fit GRB prompt emission data by presenting an example fit to the archetypical photospheric burst GRB 090902B. A detailed discussion regarding the physics of the fit is beyond the scope of this paper. The fit is included in this work as a proof of concept. 

%GRB 090902B was detected by both Fermi GBM and Fermi LAT and the burst was very bright. In a time-resolved analysis, \citet{Ryde2011} showed that the burst could be divided into two different epochs depending on the spectral shape. The spectrum during epoch 1, which lasts the first $\sim 12~$s, is very hard. In epoch two, the spectrum softens. A softening of the spectrum could be the result of dissipation below the photosphere. Here, we fit one of the time-resolved spectra from epoch 2. The fit can be seen in Figure \ref{Fig:Fit}.

%\begin{figure}
%\begin{centering}
    %\includegraphics[width=\columnwidth]{Fit_090902B_nuFnu.png}
    %\caption{A fit to GRB 090902B between $20.9~$s and $21.5~$s. }
%    \label{Fig:Fit}
%\end{centering}
%\end{figure}

\begin{figure}
\begin{centering}
    \includegraphics[width=\columnwidth]{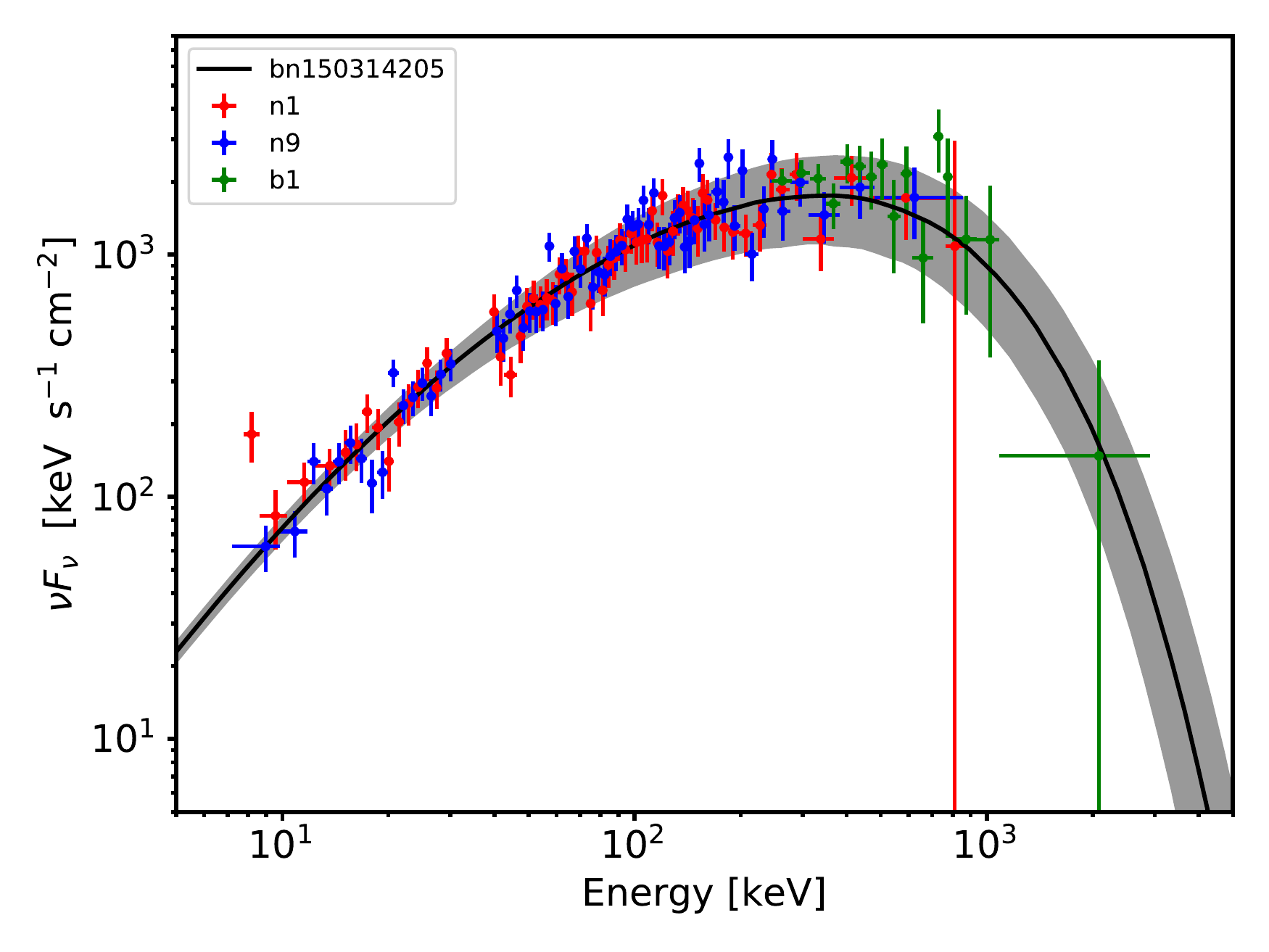}
    \includegraphics[width=\columnwidth]{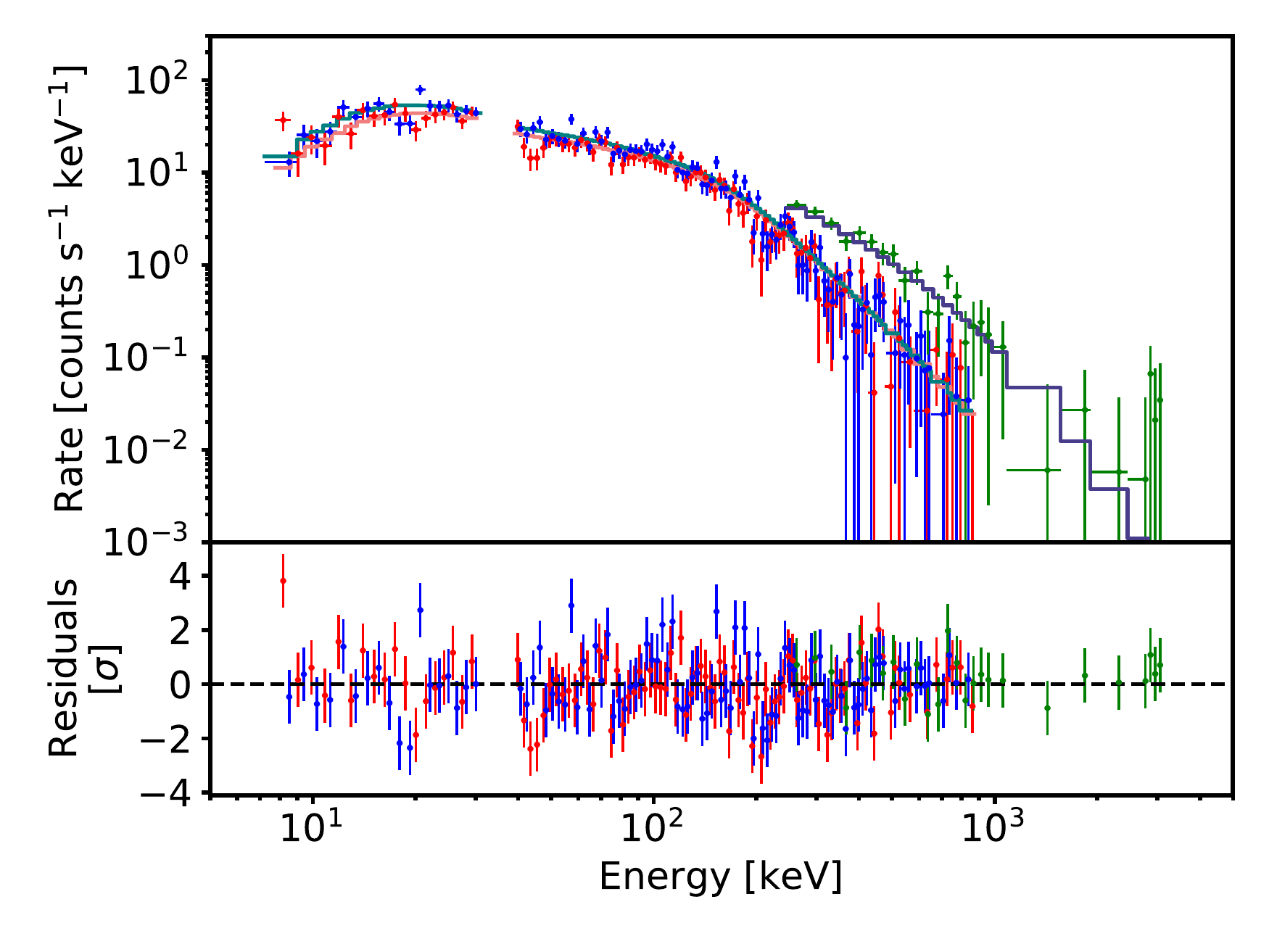}
    \caption{Time-resolved spectrum in GRB 150314A from a narrow time-bin at around 4.6 seconds after the GBM trigger. Upper panel: $\nu F_\nu$-spectrum of the best fit RMS model. Two breaks are present at around 30 and 400 keV. The best fit model is depicted by the black line and the grey region is its statistical uncertainty. The data points are derived from the counts fit, and correspond to three of the triggered GBM detectors. Note that the non-linearity of the GBM response matrix means that the data points will not be accurate in a $\nu F_\nu$-spectrum, they are here shown only for visual purpose. Lower panel: The best fit model to the observed count data, including the residuals, which show random variation, indicating a good fit.}
    \label{Fig:Fit}
\end{centering}
\end{figure}

Figure \ref{Fig:Fit} shows a fit to one of the non-thermal spectra in GRB 150314A with our RMS model. The data is from a narrow time-bin at around 4.6 seconds after the trigger. The fit indicates that there are two spectral breaks present, one at around 30 keV and the other at around 400 keV. 
The high-energy break corresponds to the energy the high-energy photons, those that reached the maximum energy $\epsilon_{\rm max} \sim 4\theta_{\rm r}$ in the RMS, have downscattered to before decoupling. Conversely, the low-energy break depends on how much the low-energy photons, those that entered the downstream with energy $\sim {\bar \epsilon}_{\rm u}$, are heated by scatterings before they reach the photosphere (see also Section \ref{sec:qualitative_predictions}).
A corresponding fit with the Band function yields $\alpha = -0.73 \pm0.06$ and a high-energy index $\beta = -2.47\pm0.25$, which are typical values for non-thermal spectra in GRBs \citep[e.g., ][]{Yu2016}. Both the Band function and the RMS model have AIC = 1610, and can therefore equally well describe the data.
The model parameters of the best fit of the RMS model are $\tau \theta = 11.3 \pm 2.9$, $R = 290\pm 50$, and $y_{\rm r} = 1.72 \pm 0.14$. The initial separation between $\theta_{\rm r}$ and $\theta_{\rm u}$ was thus relatively large and the thermalization moderate, which allow for the broad, non-thermal shape of the spectrum. The slope of the power-law segment at around 200 keV reveals that quite a lot of energy has been dissipated in the shock, causing a large value of $y_{\rm r}$. Assuming a Lorentz factor of $\Gamma = 300$, the parameters can be decoupled.\footnote{The details of how $\Gamma$ is related to the fitted parameters will be described in an upcoming paper, which focuses on GRB data analysis using the KRA model.} 
This gives $\theta_{\rm r} = 0.055$ and $\tau_{\rm i} = 206$. Translating this into the physical RMS parameters yields the upstream 4-velocity as $\beta_{\rm u}\gamma_{\rm u} = 1.89$, an upstream temperature of $\theta_{\rm u} = 8.81\times 10^{-5}$, and a photon to proton ratio of $n_\gamma/n_{\rm p} = 2.01 \times 10^5$. This fit thus illustrates that our model can be used to study the flow properties and the shock physics in observed GRBs.

%
%%%%%%%%%%%%%%%%%%%%%%%%%%%%%%%%%%%%%%
%%%%%%%%%%%%% DISCUSSION %%%%%%%%%%%%%
%%%%%%%%%%%%%%%%%%%%%%%%%%%%%%%%%%%%%%
\section{Discussion and conclusion}\label{Sec:Discussion}
Dissipation in the optically thick regions of a GRB jet has the potential to generate a wide variety of released photospheric spectral shapes, and it is, therefore, a promising candidate for the prompt emission. Although RMSs are a natural dissipation mechanism, so far, no such model has been fitted to data. In this paper, we have for the first time performed a fit to a time-resolved spectrum of the prompt emission in a GRB using an RMS model. This allowed us to determine the physical properties of the initial shock, such as its speed and the upstream photon temperature.

The main reason for the previous lack of fitted prompt spectra within an RMS framework is that RMSs are computationally expensive to simulate from first principles. To overcome this obstacle, we developed an approximate model (KRA; Kompaneets RMS approximation, see Figure \ref{Fig:Threezone} for a schematic) based on the similarities between the bulk Comptonization of photons crossing an RMS and thermal Comptonization of photons on hot electrons. By comparing the simulated spectra from \komrad, a code employing the KRA, to those generated by a special relativistic radiation hydrodynamics code, we verified that the KRA can indeed accurately reproduce the RMS and downstream spectra from the full simulations in a wide parameter range (see Figures \ref{Fig:RunAE} and \ref{Fig:RunF}). 

We connected the KRA to GRB prompt observations by creating a minimal shock model considering a single RMS occurring well below the photosphere. The downstream of the shock is allowed to thermalize and cool adiabatically as it advects to the photosphere, where its radiation is released (see Figure \ref{Fig:jet} for a schematic). The model has only three free parameters determining the shape of the released spectrum: the combined parameter $\tau\theta$ that determines the amount of thermalization, $R$ that determines the extent of the power-law segment, and $y_{\rm r}$ that determines the hardness of the power-law (see Figure \ref{Fig:Parameter_interplay}). Additionally, there are two, post-processing parameters for the normalization and the frequency-shift. We generated 125 spectra using the model and, after accounting for broadening of the observed spectrum due to high-latitude effects and a radially varying photosphere, performed a fit to a broad spectrum in a narrow time bin of GRB 150314A as a proof of concept (see Figure \ref{Fig:Fit}).

\subsection{Qualitative spectral features}\label{sec:qualitative_predictions}
Within the minimal shock model developed in this paper, there are some clear observational predictions. The spectra will consist of smooth low- and high-energy cutoffs, with a power-law segment in between. The low-energy cutoff is very smooth due to the broadening effects discussed in Section \ref{sec:broad}, while there is an exponential cutoff at the highest energies. Typically, a single-break function is used to fit the spectral GRB data, e.g., the Band function, with its peak at $E_{\rm p}$. Depending on the hardness of the power-law segment, $E_{\rm p}$ can either correspond to the low-energy cutoff (when $y_{\rm r} < 1$) or the high-energy cutoff (when $y_{\rm r} > 1$). Breaks both above and below the peak energy have been detected. Additional breaks at low energies ($< 10~$keV) were reported in, e.g., \citet{Stroh1998}, while additional high-energy breaks are discussed in, e.g., \citet{Barat1998}. Within our model, bursts that have an additional low-energy break should be well fitted with an exponential cutoff above $E_{\rm p}$ or very soft values of the high-energy power-law index $\beta$ in the Band-function. Conversely, bursts that have reported high-energy breaks above $E_{\rm p}$ should produce hard low-energy slopes, as long as the smooth curvature is within the detector energy range. However, we note that many of our generated spectra will appear as a single smooth curvature over a large range of energies (see for instance the best fit in Figure \ref{Fig:Fit}), due to the effects of thermalization on the downstream spectrum, as well as the broadening effects of high-latitude emission and radially varying emission.

As it propagates toward the jet photosphere, the downstream spectrum will tend toward a Wien spectrum at the Compton temperature. The more thermalized the downstream spectrum becomes, the more difficult it will be to retrieve the original shock parameters. Once the spectrum has relaxed into a Wien spectrum, the shock information is lost. This is an inherit degeneracy in photospheric models that is important to keep in mind when drawing conclusions about the physics from the parameter estimation.

\subsection{Optical emission}
Although the curvature of the spectrum is very smooth and although it may be outside the observable energy range of the prompt detectors, the Rayleigh-Jeans limit always exists in our spectra at low energies. Therefore, our minimal model with a single shock cannot account for low-energy observations such as optical during the prompt phase. Thus, we must conclude that any early optical emission is part of the afterglow. Early optical observations are rare, and very few are within $\sim 100~$s after trigger (see \citet{Oganesyan2021} for a recent example). Optical detections are commonly reported as prompt as long as they are observed within the $T_{90}$ of the GRB\footnote{The $T_{90}$ is defined as the time during which 90\% of the total fluence was detected, from 5\% to 95\%.} \citep{Yost2007, Klotz2009}. This definition disregards whether the GRB had a quiescent period within the active phase or not. Given that most optical observations occur quite late, they could be the onset of the afterglow. This highlights the need for early optical observations in GRBs, which have the power to discriminate between the current models of the prompt emission \citep[see also][]{Oganesyan2019}. Early optical observations also have the potential to discern if GRBs are significant contributors to the observed ultra-high-energy cosmic ray flux, as discussed in \citet{Samuelsson2019,Samuelsson2020}.

\subsection{Recollimation and multiple shocks}
Recollimation shocks below the photosphere and their connection to the prompt emission in GRBs have been investigated by several authors \citep[e.g.,][]{Gottlieb2019}. Although not discussed in this paper, we expect the KRA to be able to model recollimation shocks as well. Such shocks have different dynamics, but bulk Comptonization is still responsible for the energy dissipation, leading to the same spectral features. That is, a power-law segment with a cutoff at high energies and a Rayleigh-Jeans slope a low energies. Indeed, oblique shocks, such as recollimation shocks, can be transformed into parallel shocks with a suitable Lorentz transformation \citep{HenriksenWestbury1988}. Therefore, a recollimation shock could plausibly be responsible for the subphotospheric dissipation in a GRB whose spectra can be well fitted with the KRA. 

The minimal shock model considered here consists of a single RMS, dissipating energy over a dynamical time. It is easy to imagine a more complex jet structure with multiple shocks and turbulence. %\added{Indeed, hydrodynamical simulations of jets display multiple shocks, jet-cocoon interactions, and turbulence below the photosphere \citep[e.g.,][]{Lopez-Camara2013, Lopez-Camara2014, Gottlieb2019}.} 
However, although the dynamics below the photosphere are complicated, it is not inconceivable that the shape of a time-resolved spectrum is dominated by a single, strong dissipation event. 
The good fit to the emission in GRB 150314A shows that the current minimal shock model can plausibly explain the data. Additional model complexity should be considered only if the current model is found inadequate to explain the observations. Further investigation will tell if the current minimal model is sufficient when applied to a larger sample of GRBs.

%\subsection{Future work}
%\textcolor{red}{Remove altogether? Add something about this in introduction if it is not there yet.} \deleted{In the current paper, we have developed an approximate RMS model and managed to successfully fit it to a prompt GRB spectra. Apart from explaining the details of the model, the paper is a proof of concept. What lies ahead, which is arguably more interesting, is using the model to get insight into the underlying physics. We aim to study specific events and explain their emission coherently within the model framework. We also wish to investigate a larger sample of GRBs to see how the model fares in general. Hopefully, this can give us a deeper understanding of the shock physics and the dynamics of the obscure jet.}

\acknowledgments
We acknowledge support from the Swedish National Space Agency (196/16) and the Swedish Research Council (Vetenskapsr\aa det, 2018-03513). C.L. is supported by the Swedish National Space Board under grant number Dnr. 107/16. F.R. is supported by the G\"oran Gustafsson Foundation for Research in Natural Sciences and Medicine. This research made use of the High Energy Astrophysics Science Archive Research Center (HEASARC) Online Service at the NASA/Goddard Space Flight Center (GSFC). In particular, we thank the GBM team for providing the tools and data.

\bibliographystyle{mnras}
\bibliography{References}

%%%%%%%%%%%%%%%%%%%%%%%%%%%%%%%%%%%%%%%%%%%%
%%%%%%%%%%%%% APP:SLOW HEATING %%%%%%%%%%%%%
%%%%%%%%%%%%%%%%%%%%%%%%%%%%%%%%%%%%%%%%%%%%
\appendix

\section{Converting between the KRA parameters and the RMS parameters}
\label{sec:parameter_conversion}
%The parameters for \komrad\ are shown in Table~\ref{tab:kompaneets}, while the parameters for the full RMS simulation are shown in Table~\ref{tab:rms}. 
In this appendix, we show how to convert between the \komrad\ parameters and the corresponding RMS parameters. The \komrad\ parameters are the upstream temperature $\theta_{\rm u,K}$, the effective electron temperature in the shock zone $\theta_{\rm r}$, and the Compton $y$-parameter of the shock $y_{\rm r}$. If one has obtained a value for the parameter $\tau\theta$ through a fit, then it is not possible to decouple $\theta_{\rm r}$ and $\theta_{\rm u,K}$ without additional information about the bulk Lorentz factor of the outflow. In that case, the final parameters will be functions of the Lorentz factor. The \radshock\ parameters are the upstream temperature $\theta_{\rm u}$, the upstream velocity in the shock rest frame $\beta_{\rm u}$, and the photon to baryon density $n_\gamma/n_{\rm p}$.

In the case of negligible magnetic fields and a radiation dominated equation of state, the relativistic shock jump conditions can be written as \citep[e.g.,][]{Beloborodov2017}

\begin{align}
    \gamma_{\rm d}(1+w_{\rm d}) &= \gamma_{\rm u}(1+w_{\rm u}),\label{eq:first_jump}\\[2.3mm]
    u_{\rm d}(1+w_{\rm d}) + \frac{w_{\rm d}}{4u_{\rm d}} &= u_{\rm u}(1+w_{\rm u}) + \frac{w_{\rm u}}{4u_{\rm u}},\label{eq:second_jump} \\[2.3mm]
    u_{\rm d}\rho_{\rm d} &= u_{\rm u}\rho_{\rm u},\label{eq:third_jump}
\end{align}
where $w = 4p/\rho c^2$ is the dimensionless enthalpy, $p$ is the pressure, $\rho$ is the matter density, $u \equiv \beta\gamma$ is the four-velocity (Lorentz factors are evaluated in the shock rest frame), and subscripts u and d indicate quantities in the upstream and downstream, respectively. The ratio of pressure to density is given by

\begin{align}
   w_{\rm d} &= \frac{4{\bar \epsilon_{\rm d}} m_{\rm e} n_\gamma}{3 m_{\rm p}n_{\rm p}},\label{eq:p_rho_d} \\[2.3mm]
    w_{\rm u} &= \frac{4{\bar \epsilon_{\rm u}} m_{\rm e} n_\gamma}{3 m_{\rm p}n_{\rm p}}\label{eq:p_rho_u},
\end{align}
where $\bar{\epsilon}$ is the average photon energy measured in units of $m_{\rm e}c^2$ and $n_\gamma/n_{\rm p}$ is equal in the upstream and the downstream in the case of a photon rich shock. From \komrad, ${\bar \epsilon}_{\rm d}$ is found through numerical integration of the spectrum inside the RMS zone at the end of dissipation. (Integration of the RMS zone instead of the downstream zone assures there is no contamination from the shock formation history. The equations given here are valid for an RMS in steady state and ${\bar \epsilon}_{\rm d}$ describes the average downstream energy from once the shock is in steady state. As the average downstream energy remains constant in planar geometry, the average photon energy in the steady state RMS spectrum equals ${\bar \epsilon}_{\rm d}$.) Furthermore, ${\bar \epsilon}_{\rm u} = 3 \theta_{\rm u}$, given that the upstream is a thermalized Wien spectrum.

Part of the energy gain across an RMS is due to plasma compression across the shock, which increases the upstream energy by a factor $(\rho_{\rm d}/\rho_{\rm u})^{1/3}$ \citep{BlandfordPayneII1981}. Using Equation \eqref{eq:third_jump}, the increase can be written as $(u_{\rm u }/u_{\rm d})^{1/3}$. The KRA cannot account for compression. Therefore, in order to generate the same RMS spectrum, the codes need different upstream temperatures

\begin{equation}
\theta_{\rm u, K} = \theta_{\rm u} \left(\frac{u_{\rm u }}{u_{\rm d}}\right)^{1/3}.
\label{eq:theta_u_over_theta_u_K}
\end{equation}
The jump conditions, together with Equation \eqref{eq:theta_u_over_theta_u_K}, assures the two codes get similar lower energy cutoff and average downstream energy. An additional equation is needed, which relates the energy gain per scattering in the shock, $\Delta \epsilon/\epsilon$, between the two models. The maximum photon energy in the shock roughly equals the relative energy gain, $\epsilon_{\rm max} \sim \Delta \epsilon/\epsilon$. In the \radshock\ simulations, we empirically find that $\epsilon_{\rm max} \approx u_{\rm u}^2 \log({\bar \epsilon}_{\rm d}/{\bar \epsilon}_{\rm u})/\xi$ with a constant value of $\xi = 55$ works well across the parameter space. In \komrad, the maximum energy is given by the $\epsilon_{\rm max} = 4\theta_{\rm r}$. Therefore, we obtain

\begin{equation}\label{eq:beta_u}
	u_{\rm u}^{2} = \xi \, \frac{4\theta_{\rm r}}{\log({\bar \epsilon}_{\rm d}/{\bar \epsilon}_{\rm u})},
\end{equation}
where $\xi = 55$. 

From a \komrad\ simulation, we know $\theta_{\rm u, K}$, $\theta_{\rm r}$, and ${\bar \epsilon}_{\rm d}$. Given the equation above, the system can be solved. Numerically, one can start by guessing $u_{\rm d}$. Then $u_{\rm u}$ and $\theta_{\rm u}$ can be found from equations \eqref{eq:theta_u_over_theta_u_K} and \eqref{eq:beta_u}. With equations \eqref{eq:p_rho_d} and \eqref{eq:p_rho_u}, the only unknown left is $n_\gamma/n_{\rm p}$, which can be solved from Equation \eqref{eq:first_jump}. If the original guess of $u_{\rm d}$ was correct, Equation \eqref{eq:second_jump} should be satisfied.

If one wishes to instead go from the RMS parameters to the \komrad\ parameters, one can find $w_{\rm d}$ and $u_{\rm d}$ numerically through equations \eqref{eq:first_jump} and \eqref{eq:second_jump}, using equations \eqref{eq:p_rho_d} and \eqref{eq:p_rho_u}. Then $\theta_{\rm u,K}$ and $\theta_{\rm r}$ are found from equations \eqref{eq:theta_u_over_theta_u_K} and \eqref{eq:beta_u}. The parameter $y_{\rm r}$ can be found iteratively by requiring that the downstream energy ${\bar \epsilon}_{\rm d}$ should be equal in both models. In practice, a qualitative first guess of $y_{\rm r}$ can be made from a plot of the RMS spectrum from \radshock: by comparing the power-law slope to the spectra in the lower panel of Figure \ref{Fig:Parameter_interplay}, the value of $y_{\rm r}$ can be estimated.

\section{When can the KRA model internal shocks?}
\label{sec:are_is_relativistic}

Consider a part of the jet that consists mainly of two masses: a slower and a faster mass with lab frame Lorentz factors $\Gamma_1 \gg 1$ and $\Gamma_2 \gg \Gamma_1$, respectively. The masses are assumed to be initially separated by a lab frame distance $\delta r \lesssim \delta l_1, \delta l_2$, where $\delta l_1$ and $\delta l_2$ are the corresponding initial lab frame widths of the slower and faster mass, respectively. The faster mass catches up to the slower mass at radius $R_{\rm i} \approx 2 \Gamma_1^2 \delta r$. By then, the plasma between the masses has been highly compressed, increasing its pressure adiabatically until a forward and a reverse shock forms. The forward and reverse shocks propagate into the slower and faster masses, respectively. The speed of the shocked region (i.e., the shared downstream, which is bounded by the forward and reverse shocks) is found by balancing the momentum flux in the rest frame of the shocked region, $\beta^2\Gamma^2 h + p$, from both sides. Here, $h \equiv \rho c^2 + e + p$ is the specific enthalpy, where $\rho$ is the mass density and $p$ is the pressure, all of which are measured in the respective rest frames of the unshocked masses. If we suppose that the initial pressure inside the two masses is small (such that $h \approx \rho c^2$), then we can solve for the lab frame Lorentz factor $\Gamma$ of the shocked material as

\begin{equation}
\Gamma^2 \approx \Gamma_1\Gamma_2 \frac{\Gamma_1 \rho_1^{1/2} + \Gamma_2 \rho_2^{1/2}}{\Gamma_1 \rho_2^{1/2} + \Gamma_2 \rho_1^{1/2}},
\label{eq:Gamma}
\end{equation}

\noindent where $\rho_1$ and $\rho_2$ are the proper densities of the respective mass, before being shocked. In a wide range of density ratios, $(\Gamma_2/\Gamma_1)^2 \gg \rho_2/\rho_1 \gg (\Gamma_1/\Gamma_2)^2$ holds and the above expression can be simplified to

\begin{equation}
\Gamma^2 \approx \Gamma_1\Gamma_2\left(\frac{\rho_2}{\rho_1}\right)^{1/2}.
\label{eq:downstream_Gamma}
\end{equation}

\noindent The condition $(\Gamma_2/\Gamma_1)^2 \gg \rho_2/\rho_1 \gg (\Gamma_1/\Gamma_2)^2$ also ensures that $\Gamma_2 \gg \Gamma \gg \Gamma_1$. The radii where the two shocks have crossed their respective masses are then $R_2 \approx 2 \Gamma^2 \delta l_2$ and $R_1 \approx 2 \Gamma_1^2 \delta l_1$ respectively. The masses are related to their widths and densities by $\delta m \approx 4 \pi r^2 \Gamma\rho\delta l$, and so the ratio of the radii where the reverse and forward shocks have crossed the respective masses can be written as

\begin{equation}
\frac{R_2}{R_1} \approx \left(\frac{\rho_1}{\rho_2}\right)^{1/2} \frac{\delta m_2}{\delta m_1}.
\label{eq:R_ratio}
\end{equation}

The relative Lorentz factors between the upstream (moving with Lorentz factor $\Gamma_1$ or $\Gamma_2$ for the forward and reverse shocks, respectively) and the downstream (Lorentz factor $\Gamma$) gives a measure of how relativistic the two shocks are. They can be computed using Equation~\eqref{eq:downstream_Gamma},

\begin{equation}
\bar{\Gamma}_2 \approx \frac{\Gamma_2}{2\Gamma} \approx \frac{1}{2}\left(\frac{\Gamma_2}{\Gamma_1}\right)^{1/2} \left(\frac{\rho_1}{\rho_2}\right)^{1/4},
\label{eq:bar_Gamma_2}
\end{equation}

\noindent and

\begin{equation}
\bar{\Gamma}_1 \approx \frac{\Gamma}{2\Gamma_1} \approx \frac{1}{2}\left(\frac{\Gamma_2}{\Gamma_1}\right)^{1/2} \left(\frac{\rho_2}{\rho_1}\right)^{1/4},
\label{eq:bar_Gamma_1}
\end{equation}

\noindent and their ratio is

\begin{equation}
\frac{\bar{\Gamma}_2}{\bar{\Gamma}_1} \approx \left(\frac{\rho_1}{\rho_2}\right)^{1/2}.
\end{equation}

\noindent The energy dissipated for each mass (in the rest frame of the shocked plasma) is $E \approx (\bar{\Gamma} - 1) \delta m c^2$, where $\bar{\Gamma}$ is the relative Lorentz factor between the up- and downstream, so that

\begin{equation}
\frac{E_2}{E_1} \approx \left(\frac{\rho_1}{\rho_2}\right)^{1/2} \frac{\delta m_2}{\delta m_1} \approx \frac{R_2}{R_1}.
\end{equation}

Based on the analysis above, we see that the collision of two masses with similar properties, $\delta m_1 \sim \delta m_2$ and $\rho_1 \sim \rho_2$ results in shocks of similar strengths, $\bar{\Gamma}_1 \sim \bar{\Gamma}_2$. The shocks also dissipate roughly the same amount of energy, $E_1 \sim E_2$, and finish dissipating roughly at the same time, $R_1 \sim R_2$. The heated radiation is located in the shared downstream between the two shocks. Since the shocks have similar strengths, and the heated radiation from both shocks sits inside plasma that propagates with the same Lorentz factor (here called simply $\Gamma$), modeling of only one shock is necessary. %Scenarios with more complex shock dynamics can of course be imagined.

The KRA can accurately model shocks as long as the relative energy gain per scattering is small less than unity, $\Delta\epsilon/\epsilon \lesssim 1$. In Appendix~\ref{sec:parameter_conversion} we found that $\Delta\epsilon/\epsilon \approx 0.018 (\gamma_{\rm u}\beta_{\rm u})^2 \log({\bar \epsilon}_{\rm d}/{\bar \epsilon}_{\rm u})$, which means the approximation is valid up to $\beta_{\rm u}\gamma_{\rm u} \sim 3\textrm{--}4$. Such a scenario is shown in Figure \ref{Fig:RunF}. For two blobs with $\rho_1 \sim \rho_2$, $\gamma_{\rm u} \beta_{\rm u} = 3$ translates to $\bar{\Gamma} \approx 3$. Using Equations~\eqref{eq:bar_Gamma_2} and \eqref{eq:bar_Gamma_1}, we find

\begin{equation}
    \frac{\Gamma_2}{\Gamma_1} \lesssim 36.
\end{equation}

\noindent As an example, two masses of similar properties that propagate with initial Lorenz factors of $\Gamma_1 \approx 50$ and $\Gamma_2 \approx 1000$ would give rise to two RMSs with $\Delta\epsilon/\epsilon < 1$, which can be modelled by the KRA.

\end{document}